\documentclass[preprint,12pt]{elsarticle}

\usepackage{amssymb}
\usepackage{amsmath}

\journal{Nuclear Physics B}

\usepackage{amsmath,amssymb,amsfonts,amsthm}

\theoremstyle{definition}
\newtheorem{definition}{Definition}[section]

\theoremstyle{plain}
\newtheorem{theorem}[definition]{Theorem}
\newtheorem{lemma}[definition]{Lemma}
\newtheorem{corollary}[definition]{Corollary}

\theoremstyle{remark}
\newtheorem{remark}[definition]{Remark}

\usepackage{graphicx}

\usepackage{tikz}
\usetikzlibrary{arrows.meta,calc,positioning,decorations.markings,decorations.pathreplacing,shapes.geometric,backgrounds,fit}

\tikzset{
  vertex/.style={circle,draw,inner sep=0pt,minimum size=5pt},
  square/.style={draw,regular polygon,regular polygon sides=4,inner sep=0pt,minimum size=0.225cm},
  triangle/.style={draw,regular polygon,regular polygon sides=3,inner sep=0pt,minimum size=0.3cm}
}

\tikzset{
  every node/.style={font=\scriptsize},
  conn/.style={circle,draw,fill=gray!25,inner sep=1.4pt,minimum size=10pt},
  pnode/.style={circle,draw,fill=white,inner sep=1.2pt,minimum size=9pt},
  qnode/.style={circle,draw,fill=gray!15,inner sep=1.0pt,minimum size=8pt},
  pathedge/.style={line width=1.1pt},
  labA/.style={font=\scriptsize, yshift=2.6mm},
  labB/.style={font=\scriptsize, yshift=-2.8mm}
}

\DeclareMathOperator{\tc}{tc}
\DeclareMathOperator{\tw}{tw}
\DeclareMathOperator{\cw}{cw}
\DeclareMathOperator{\td}{td}
\DeclareMathOperator{\nd}{nd}
\DeclareMathOperator{\pw}{pw}
\DeclareMathOperator{\diam}{diam}
\DeclareMathOperator{\dist}{dist}
\DeclareMathOperator{\vc}{vc}
\DeclareMathOperator{\fvs}{fvs}
\DeclareMathOperator{\vi}{vi}
\DeclareMathOperator{\mw}{mw}
\DeclareMathOperator{\cvd}{cvd}

\newcommand{\proofpara}[1]{\medskip\noindent\textit{#1.}\ }

\usepackage[hidelinks]{hyperref}

\usepackage{xcolor}

\begin{document}

\begin{frontmatter}



\title{On the Parameterized Complexity of $s$-Club Cluster Edge Deletion}


\author{Ajinkya Gaikwad} 

\affiliation{
  organization={Indian Institute of Science Education and Research Pune},
  addressline={Dr. Homi Bhabha Road, Pashan},
  city={Pune},
  postcode={411008},
  state={Maharashtra},
  country={India}
}
\begin{abstract}
We study the parameterized and kernelization complexity of the \emph{\textsc{$s$-Club Cluster Edge Deletion}} problem, a natural distance-bounded generalization of \emph{\textsc{Cluster Edge Deletion}}.  
Given a graph \(G=(V,E)\) and integers \(k,s\), the goal is to delete at most \(k\) edges so that every connected component in the resulting graph has diameter at most \(s\).  
 This problem captures a broad class of distance-constrained graph modification problems that interpolate between clustering and connectivity control.
On the structural side, we settle an open question of Montecchiani, Ortali, Piselli, and Tappini (\emph{Theoretical Computer Science}, 2023) by proving that the problem is W[1]-hard when parameterized by \emph{pathwidth} plus the maximum number of allowed $s$-clubs, and consequently also by \emph{treewidth} plus the maximum number of allowed $s$-clubs, showing that the diameter bound \(s\) is indispensable for fixed-parameter tractability.
In contrast, we identify several width and density parameters for which the dependence on \(s\) is unnecessary: the problem is fixed-parameter tractable when parameterized by \emph{treedepth}, \emph{neighborhood diversity}, or the \emph{cluster vertex deletion number}.  
This generalizes previous FPT results known only for \(s=1\).  
We also complement these results by proving that no polynomial kernel exists when parameterized by the \emph{vertex cover number}, even for \(s=2\).
Finally, we present an FPT bicriteria approximation scheme that, for graphs excluding long induced cycles, runs in time \(f(k,1/\epsilon)\cdot n^{\mathcal{O}(1)}\) and produces a solution of size at most \(k\) whose components have diameter at most \((1+\epsilon)s\).   
We also initiate the study of the directed variant, \textsc{$s$-Club Cluster Arc Deletion}, and show that it is W[1]-hard when parameterized by \(k\), even on directed acyclic graphs.  
\end{abstract}




\end{frontmatter}



\section{Introduction}

Graph modification problems lie at the heart of algorithmic graph theory.  
They ask whether a given graph can be transformed into one satisfying a desired structural property 
by performing a limited number of edge or vertex edits.  
This framework has proved remarkably robust: by varying the target property, 
one captures many classical problems in clustering, network design, and computational biology.  
Among these, perhaps the most fundamental is the \emph{clique-based} clustering model, 
which seeks to delete a few edges so that the resulting graph is a disjoint union of cliques.  
This gives rise to the well-studied \textsc{Cluster Edge Deletion} problem.

\medskip
\noindent
A natural generalization replaces cliques by \emph{$s$-clubs}: 
vertex sets whose induced subgraphs have diameter at most $s$.  
An $s$-club can be viewed as a relaxed clique that allows distances up to $s$, 
thus tolerating missing edges while maintaining local structure.  
The corresponding modification problem, called \textsc{$s$-Club Cluster Edge Deletion},  
asks whether at most $k$ edges can be deleted from a graph $G=(V,E)$ 
so that every connected component of the resulting graph has diameter at most $s$.  
For $s=1$ this coincides with \textsc{Cluster Edge Deletion},  
while larger values of $s$ model progressively looser notions of cluster cohesion.

\medskip
\noindent
\textbf{Other $s$-club modification problems.}
The $s$-club concept has inspired a wide family of related graph modification problems, 
each differing in the allowed operations but sharing the common goal of enforcing $s$-club structure.  
Examples include \textsc{$s$-Club Cluster Vertex Deletion}, where vertices are deleted, 
\textsc{$s$-Club Cluster Edge Addition}, where edges are added, 
and \textsc{$s$-Club Cluster Editing}, where both edge additions and edge deletions are allowed.  
The \emph{vertex-deletion} variant was recently investigated by 
Chakrabortty, Chandran, and Pillai~\cite{10.1007/978-3-030-79987-8_11},  
who proved that for every $s \ge 2$ the problem is NP-hard even on bounded-degree planar bipartite graphs 
and APX-hard on bounded-degree bipartite graphs, 
while being polynomial-time solvable on trapezoid graphs.  
The \emph{editing} variant, where both edge additions and edge deletions are allowed, and the \emph{addition} variants were explored by 
Figiel, Himmel, Nichterlein, and Niedermeier~\cite{Figiel2021} 
and by Bazgan, Heggernes, Nichterlein, and Pontoizeau~\cite{BAZGAN2025247}, respectively.  
Figiel et al.\ showed that \textsc{2-Club Cluster Editing} is W[2]-hard with respect to the number of edge modifications, 
and that \textsc{2-Club Cluster Vertex Deletion} is fixed-parameter tractable 
but admits no polynomial kernel under standard complexity assumptions.  
Bazgan et al.\ proved that the \textsc{$s$-Club Edge Addition} problem is W[2]-hard 
with respect to the number of edges to add, even when restricted to split graphs, 
and that minimizing the number of edges retained to maintain diameter $2$ or $3$ is NP-complete.  
Together, these studies demonstrate that permitting additions or combined editing 
operations significantly increases computational hardness compared to the pure deletion setting.

\medskip
\noindent
\textbf{Previous work.}
The case $s=1$, corresponding to the classical \textsc{Cluster Edge Deletion} problem, 
has been extensively investigated in both classical and parameterized complexity.  
It remains NP-hard even on highly restricted graph classes such as planar graphs~\cite{golovach_et_al:LIPIcs.SWAT.2018.23}, 
$C_4$-free graphs of maximum degree four~\cite{KOMUSIEWICZ20122259}, and various hereditary graph classes.  
In particular, Gao et al.~\cite{GAO20132763} proved that \textsc{Cluster Deletion} is NP-hard on 
$(C_5, P_5, \text{bull}, \text{fork}, \text{co-gem},$ $ \text{4-pan}, \text{co-4-pan})$-free graphs 
as well as on $(2K_2, 3K_1)$-free graphs. 
From a classical viewpoint, the problem is polynomial-time solvable on split graphs for $s=1$, 
and the case $s=3$ is trivially easy since every connected split graph has diameter at most three.  
The complexity of the intermediate case $s=2$, however, was open. On the positive side, the problem for $s=1$ is polynomial-time solvable on interval graphs~\cite{konstantinidis_et_al:LIPIcs.MFCS.2019.12}.

A long line of research has focused on the parameterized complexity of the problem.  
When parameterized by the number of edge deletions $k$, 
\textsc{Cluster Edge Deletion} admits a fixed-parameter algorithm 
running in time $\mathcal{O}^{*}(1.404^{k})$~\cite{TSUR2022106171}, 
which currently represents the best known bound for this parameterization. 
Beyond the solution-size parameter, Italiano et al.~\cite{10.1007/978-3-031-27051-2_31} investigated the problem under 
structural measures such as \emph{twin cover} and \emph{neighborhood diversity}, 
for which they designed fixed-parameter algorithms, thereby identifying broader classes of graphs 
on which \textsc{Cluster Edge Deletion} remains tractable.  
Also, Komusiewicz and Uhlmann~\cite{10.1007/978-3-642-18381-2_29} established an FPT algorithm 
for the problem parameterized by the \emph{cluster vertex deletion number}.
For $s=2$, Abu-Khzam et al. provided an FPT with a running time $\mathcal{O}^{*}(2.692^{k})$.

However, once $s>1$, the overall complexity landscape remains far less understood, leaving a number of natural questions about the role of $s$ in determining tractability.
A recent result of Montecchiani et al.~\cite{MONTECCHIANI2023114041}  
established that the problem is fixed-parameter tractable when parameterized by $s + \mathrm{tw}(G)$,  
where $\mathrm{tw}(G)$ denotes the treewidth of the input graph,  
and explicitly asked whether dependence on $s$ is necessary—that is,  
whether \textsc{$s$-Club Cluster Edge Deletion} remains FPT under treewidth alone.

\medskip
\noindent
\textbf{Our results.}
We revisit \textsc{$s$-Club Cluster Edge Deletion} through the lens of parameterized and structural complexity,
with the goal of delineating the precise boundary of its tractability. 
Our contributions unify and extend several existing lines of research.

\begin{itemize}

\item \emph{Parameterized intractability.}
We settle an open problem posed by Montecchiani, Ortali, Piselli, and Tappini
(\emph{Information and Computation}, 2023) by proving that the problem is W[1]-hard
when parameterized by \emph{pathwidth}—and consequently by \emph{treewidth}.
In fact, our reduction yields a stronger result: the problem remains W[1]-hard
even when parameterized by the combined parameter \(\pw(G) + d\),
where d denotes the maximum number of s-clubs allowed in the modified graph.
This firmly establishes that the diameter bound s is essential for fixed-parameter tractability.

\item \emph{Structural parameterizations.}
We identify several graph parameters for which the dependence on \(s\) is unnecessary:
the problem is fixed-parameter tractable when parameterized by
\emph{treedepth}, \emph{neighborhood diversity}, or \emph{cluster vertex deletion number}.
These results extend to all \(s\ge1\) the algorithms of Italiano, Konstantinidis, and Papadopoulos
(\emph{Algorithmica}, 2023) for the case \(s=1\),
and generalize the FPT algorithm of Komusiewicz and Uhlmann
(\emph{SOFSEM}, 2011) for the cluster vertex deletion parameter.

\item \emph{Solution-size parameterization.}
While \textsc{$s$-Club Cluster Edge Deletion} is known to be FPT under the combined parameter \(s+k\),
the case of parameter \(k\) alone remains unresolved.
We make progress by designing an FPT bicriteria approximation scheme that, 
for graphs excluding long induced cycles, runs in time \(f(k,\varepsilon)\,n^{\mathcal{O}(1)}\)
and outputs a deletion set of size at most \(k\)
yielding components of diameter at most \((1+\varepsilon)s\).

\item \emph{Directed generalization.}
We introduce and analyze \textsc{$s$-Club Cluster Arc Deletion} on directed graphs.
Given a digraph \(D=(V,A)\), the goal is to delete at most \(k\) arcs so that in \(D-F\)
every weakly connected component \(C\) satisfies:
for all \(u,v\in V(C)\), if \(u\) is reachable from \(v\), then there exists a directed path from \(v\) to \(u\)
of length at most \(s\).
This directed formulation extends the undirected notion by bounding asymmetric reachability distances.
We show that this variant is W[1]-hard when parameterized by the solution size \(k\),
even when restricted to directed acyclic graphs (DAGs).
\end{itemize}

\section{Preliminaries}

\subsection{Basic Graph Notation}
We consider finite, simple, undirected, and directed graphs.  
For a graph \(G\), its vertex and edge sets are denoted by \(V(G)\) and \(E(G)\), respectively.  
An undirected edge between vertices \(u\) and \(v\) is written as \(\{u,v\}\), while a directed edge (or \emph{arc}) from \(u\) to \(v\) is written as \((u,v)\).  
For a vertex \(v\in V(G)\), the \emph{open neighborhood} of \(v\) is 
\(N_G(v):=\{u\in V(G)\mid \{u,v\}\in E(G)\}\),  
and its \emph{closed neighborhood} is \(N_G[v]:=N_G(v)\cup\{v\}\).  
The \emph{degree} of \(v\) is \(\deg_G(v):=|N_G(v)|\).

\smallskip
For a directed graph \(D=(V,A)\), the \emph{out-neighborhood} and \emph{in-neighborhood} of a vertex \(v\) are defined as  
\(N_D^{+}(v):=\{u\mid (v,u)\in A\}\) and \(N_D^{-}(v):=\{u\mid (u,v)\in A\}\).  
The corresponding \emph{outdegree} and \emph{indegree} are  
\(\deg_D^{+}(v):=|N_D^{+}(v)|\) and \(\deg_D^{-}(v):=|N_D^{-}(v)|\), respectively.  
A \emph{path} in an undirected or directed graph is a sequence of distinct vertices  
\(v_1,v_2,\ldots,v_\ell\) such that consecutive vertices are adjacent (or properly oriented in the directed case).  
The \emph{length} of a path is \(\ell-1\).  
For two vertices \(u,v\in V(G)\), the \emph{distance} \(\dist_G(u,v)\) is the length of a shortest path between \(u\) and \(v\),  
and is defined to be \(\infty\) if no path exists.  
The \emph{diameter} of a graph \(G\) is the maximum distance between any two vertices in the same connected component, that is,  
\[
\diam(G):=\max_{u,v\in V(G)}\dist_G(u,v).
\]
For a vertex subset \(S\subseteq V(G)\), we write \(G[S]\) to denote the subgraph of \(G\) induced by \(S\).  

\smallskip
An \emph{orientation} of an undirected graph \(G=(V,E)\) is a directed graph \(\Lambda=(V,A)\) obtained by replacing each edge \(\{u,v\}\in E(G)\) with exactly one of the arcs \((u,v)\) or \((v,u)\).  
Given an edge-weight function \(w:E(G)\to\mathbb{Z}_{\ge1}\), the \emph{weighted outdegree} of a vertex \(u\) under the orientation \(\Lambda\) is defined as  
\[
w_{\mathrm{out}}(u):=\sum_{(u,v)\in A(\Lambda)}w(\{u,v\}),
\]
which coincides with the usual outdegree when \(w(e)=1\) for all \(e\in E(G)\).  
All logarithms are taken to base~2 unless stated otherwise.  
We use standard asymptotic notation, and \(\mathcal{O}^{*}(\cdot)\) suppresses polynomial factors in~\(n\).

\subsection{Structural Graph Parameters} 

We now review the concept of a tree decomposition introduced by Robertson and Seymour \cite{Neil}.
Treewidth is a  measure of how “tree-like” the graph is.
\begin{definition}[Robertson and Seymour \cite{Neil}]  A {\it tree decomposition} of a 
graph~$G=(V,E)$  is a tree $T$ together with a 
collection of subsets $X_t$ (called \emph{bags}) of $V$ labeled by the nodes $t$ of $T$ such that 
$\bigcup_{t\in T}X_t=V $ and (1) and (2) below hold:
\begin{enumerate}
			\item For every edge $(u,v) \in E(G)$, there  is some $t$ such that $\{u,v\}\subseteq X_t$.
			\item  (Interpolation Property) If $t$ is a node on the unique path in $T$ from $t_1$ to $t_2$, then 
			$X_{t_1}\cap X_{t_2}\subseteq X_t$.
		\end{enumerate}
	\end{definition}

\begin{definition}[Robertson and Seymour \cite{Neil}]\label{deftw} The {\it width} of a tree decomposition is
the maximum value of $|X_t|-1 $ taken over all the nodes $t$ of the tree $T$ of the decomposition.
The \emph{treewidth} $\tw(G)$ of a graph $G$  is the  minimum width among all possible tree decompositions of $G$.
\end{definition} 

\begin{definition}\label{defpw}\rm 
    If the tree $T$ of a tree decomposition is a path, then we say that the tree decomposition 
    is a {\it path decomposition}. The \emph{pathwidth}  ${\pw}(G)$ of a graph $G$  is the  minimum width among all possible path decompositions of $G$.
\end{definition}

\noindent A rooted forest is a disjoint union of rooted trees. Given a rooted forest $F$, its \emph{transitive closure} is a graph $H$ in which $V(H)$ contains all the nodes of the rooted forest, and $E(H)$ contain an edge between two vertices whenever those two vertices form an ancestor-descendant pair in the forest $F$.

   \begin{definition}\label{deftd}
        {\rm  The {\it treedepth} of a graph $G$ is the minimum height of a rooted forest~$F$ whose transitive closure contains the graph $G$ as a subgraph. It is denoted by $\td(G)$.}
    \end{definition}

\begin{definition}\label{defvc}\rm 
A set $S \subseteq V(G)$ is a \emph{vertex cover} of $G=(V,E)$ if  each edge in $E$ has at least one  endpoint in $S$.   The \emph{size} of a smallest vertex cover of  $G$ is the \emph{vertex cover number} of $G$ and is denoted as $\vc(G)$.
\end{definition}

\noindent  Ganian \cite{dmtcs:2136} introduced
a new parameter called twin-cover and showed that it is capable of solving
a wide range of hard problems. 
According to Ganian \cite{dmtcs:2136}, ``The twin-cover is a direct generalization of vertex cover towards richer graph classes,
including dense graph classes (which have unbounded treewidth)."
 Here we relax the definition of vertex cover so that not all edges need to be covered.
\begin{definition}[Ganian \cite{dmtcs:2136}]
An edge $(u,v)\in E(G)$ is a \emph{twin edge} of $G$ if $N[u]=N[v]$. 
\end{definition}

\begin{definition}[Ganian \cite{dmtcs:2136}]\label{deftc}
  A set $X\subseteq V(G)$ is a \emph{twin-cover} of $G$ if every edge in~$G$ is either twin edge or incident to a vertex in $X$. We then say that $G$ has \emph{twin-cover number}~$k$ if $k$ is the minimum possible size
of a twin-cover of $G$. It is denoted as $\tc(G)$
\end{definition}

Two different vertices $u$, $v$ are called \emph{true twins} if $N[u]=N[v]$. 
Likewise, $u$, $v$ are called  \emph{false twins} if  $N(u)=N(v)$.
In general, $u$, $v$ are called  \emph{twins} if they are either true twins or false twins.
  If they are twins, we say that they have the same \emph{neighbourhood type}.

\begin{definition}[Lampis \cite{Lampis}]\label{defnd}
A graph  $G=(V,E)$
 has \emph{neighbourhood diversity} at most $d$, if there exists a partition of $V$
 into at most $d$
 sets (we call these sets {\it type classes}) such that all the vertices in each set have the same neighbourhood type. It is denoted as $\nd(G)$.
\end{definition}
 
\begin{definition}\label{deffvs}\rm 
 A \emph{feedback vertex set}  of a graph  $G$ is a set of vertices whose removal turns $G$ into a forest. The minimum size of a feedback vertex set in $G$ is the {\it feedback vertex set number} of $G$, denoted by ${\fvs}(G)$.
\end{definition}

\begin{definition}\label{cvd}\rm
    The \emph{cluster vertex deletion number} of a graph is the minimum number of its vertices whose deletion results in a disjoint union of complete graphs and it is denoted as $\cvd(G)$.
\end{definition}

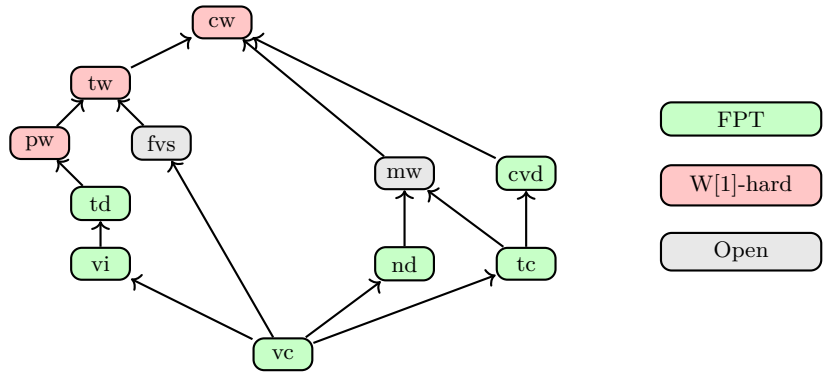
\begin{figure}[t]
\centering
\begin{tikzpicture}[
    auto, scale=.8,
    param/.style={
      rectangle, rounded corners, thick, draw=black,
      align=center, minimum width=2.2em, minimum height=1.2em,
      font=\footnotesize
    },
    fpt/.style={param, fill=green!22},
    hard/.style={param, fill=red!22},
    open/.style={param, fill=gray!18}
]
\node[fpt]  (vc)  at (0,0)        {$\vc$};
\node[fpt]  (nd)  at (2,1.5)      {$\nd$};
\node[fpt]  (tc)  at (4,1.5)      {$\tc$};
\node[fpt]  (vi)  at (-3,1.5)     {$\vi$};
\node[fpt]  (td)  at (-3,2.5)     {$\td$};
\node[open] (fvs) at (-2,3.5)     {$\fvs$};
\node[hard] (pw)  at (-4,3.5)     {$\pw$};
\node[hard] (tw)  at (-3,4.5)     {$\tw$};
\node[open] (mw)  at (2,3)        {$\mw$};
\node[fpt]  (cvd) at (4,3)        {$\cvd$};
\node[hard] (cw)  at (-1,5.5)     {$\cw$};

\draw[->, thick] (vc)--(vi);
\draw[->, thick] (vc)--(nd);
\draw[->, thick] (vc)--(tc);
\draw[->, thick] (vi)--(td);
\draw[->, thick] (td)--(pw);
\draw[->, thick] (vc)--(fvs);
\draw[->, thick] (nd)--(mw);
\draw[->, thick] (tc)--(mw);
\draw[->, thick] (tc)--(cvd);
\draw[->, thick] (pw)--(tw);
\draw[->, thick] (fvs)--(tw);
\draw[->, thick] (tw)--(cw);
\draw[->, thick] (mw)--(cw);
\draw[->, thick] (cvd)--(cw);

\node[param, fill=green!22, minimum width=6em, anchor=west] (legf) at (6.2,3.9) {FPT};
\node[param, fill=red!22,   minimum width=6em, anchor=west] (legh) at (6.2,2.8) {W[1]-hard};
\node[param, fill=gray!18,  minimum width=6em, anchor=west] (lego) at (6.2,1.7) {Open};

\end{tikzpicture}

\caption[{Parameterized complexity landscape for the problem.}]{
Relationship between graph parameters and our results for 
\textsc{$s$-Club Cluster Edge Deletion}. 
An arrow \(A \to B\) indicates that there exists a function \(f\) such that 
\(f(A(G)) \ge B(G)\) for all graphs \(G\).
Color coding: parameters in green correspond to those for which the problem is fixed-parameter tractable (FPT), including vertex cover (\(\vc\)), neighborhood diversity (\(\nd\)), twin cover (\(\tc\)), vertex integrity (\(\vi\)), cluster vertex deletion number (\(\cvd\)), and treedepth (\(\td\)). 
Parameters in red indicate W[1]-hardness, namely pathwidth (\(\pw\)), treewidth (\(\tw\)), and cliquewidth (\(\cw\)). 
Finally, parameters in gray correspond to cases that remain open, specifically the feedback vertex set and modular width.
}
\label{overview}
\end{figure}

\vspace{1ex}
The relationships between these parameters and the parameterized complexity of our problem are summarized in Figure~\ref{overview}.

\subsection{Problems Used in Reductions}\label{problems section}

In this subsection, we define the problems used as sources in our reductions.

Our main hardness reduction uses the \textsc{Minimum Maximum Outdegree} problem, 
which asks whether the edges of a weighted graph can be oriented so that the weighted 
outdegree of every vertex remains bounded by a prescribed threshold.

\vspace{3mm}

\noindent\fbox{%
  \begin{minipage}{32.4em}\label{MMO}
    {\sc Minimum Maximum Outdegree}\\
    \noindent{\bf Input:} An undirected graph $G=(V,E)$, an edge weighting 
    $w:E\to\mathbb{Z}_{\ge 1}$ given in unary, and a positive integer $r$. \\
    \noindent{\bf Question:} Does there exist an orientation $\Lambda$ of $G$ such that 
    $w_{\mathrm{out}}(u)\le r$ for every $u\in V(G)$?
  \end{minipage}%
}

\vspace{2mm}

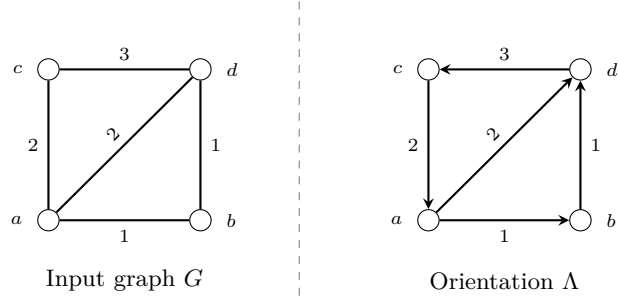
\begin{figure}[ht]
\centering
\begin{tikzpicture}[scale=1.0,
  vertex/.style={circle,draw,inner sep=1.5pt,minimum size=8pt,fill=white},
  edge/.style={thick},
  orient/.style={thick,->,>=stealth},
  font=\small
]

\node[vertex] (a) at (0,0) {};
\node[vertex] (b) at (2,0) {};
\node[vertex] (c) at (0,2) {};
\node[vertex] (d) at (2,2) {};

\node[left=2pt of a] {$a$};
\node[right=2pt of b] {$b$};
\node[left=2pt of c] {$c$};
\node[right=2pt of d] {$d$};

\draw[edge] (a)--(b) node[midway,below] {1};
\draw[edge] (a)--(c) node[midway,left] {2};
\draw[edge] (b)--(d) node[midway,right] {1};
\draw[edge] (c)--(d) node[midway,above] {3};
\draw[edge] (a)--(d) node[midway,sloped,above] {2};

\node[draw=none,font=\small] at (1,-0.8) {Input graph $G$};

\node[vertex] (a2) at (5,0) {};
\node[vertex] (b2) at (7,0) {};
\node[vertex] (c2) at (5,2) {};
\node[vertex] (d2) at (7,2) {};

\node[left=2pt of a2] {$a$};
\node[right=2pt of b2] {$b$};
\node[left=2pt of c2] {$c$};
\node[right=2pt of d2] {$d$};

\draw[orient] (a2)--(b2) node[midway,below] {1};
\draw[orient] (c2)--(a2) node[midway,left] {2};
\draw[orient] (b2)--(d2) node[midway,right] {1};
\draw[orient] (d2)--(c2) node[midway,above] {3};
\draw[orient] (a2)--(d2) node[midway,sloped,above] {2};

\node[draw=none,font=\small] at (6,-0.8) {Orientation $\Lambda$};

\draw[dashed,gray] (3.3,-1) -- (3.3,3);

\end{tikzpicture}
\caption{Illustration of \textsc{Minimum Maximum Outdegree} with $r=3$: the input weighted graph $G$ (left) and an orientation $\Lambda$ where every vertex has weighted outdegree at most 3 (right).}
\label{fig:MMO-example}
\end{figure}

An instance of \textsc{Minimum Maximum Outdegree} consists of an undirected graph 
$G=(V,E)$ together with an edge-weight function 
$w:E\to\mathbb{Z}_{\ge 1}$ and a positive integer $r$. 
The task is to determine whether there exists an orientation $\Lambda$ of $G$ 
such that the weighted outdegree of every vertex is at most $r$, where

\[
w_{\mathrm{out}}(u):=\sum_{(u,v)\in A(\Lambda)} w(\{u,v\}).
\]
Figure~\ref{fig:MMO-example} illustrates an instance of 
\textsc{Minimum Maximum Outdegree} together with a feasible orientation.
The problem is known to be W[1]-hard when parameterized by the vertex cover number of the input graph~\cite{bodlaender2022problemshardtreewidtheasy}.

\vspace{4mm}

Another source problem used in our reductions is the classical 
\textsc{Red--Blue Dominating Set} problem.

\vspace{3mm}

\noindent\fbox{%
  \begin{minipage}{32.4em}
    {\sc Red--Blue Dominating Set}\\
    \noindent{\bf Input:} A bipartite graph 
    $G=(R\cup B,E)$ with red vertices $R$ and blue vertices $B$, 
    and an integer $k$. \\
    \noindent{\bf Question:} Does there exist a subset 
    $D\subseteq R$ of size at most $k$ such that every vertex of $B$ 
    has a neighbor in $D$?
  \end{minipage}%
}

\vspace{2mm}

In the \textsc{Red--Blue Dominating Set} problem, the goal is to select at most 
$k$ red vertices so that every blue vertex has at least one selected red neighbor. 
The problem is parameterized by $|B|$.
It is known that \textsc{Red--Blue Dominating Set}, parameterized by $|B|$, 
does not admit a polynomial kernel unless 
$\mathrm{NP}\subseteq \mathrm{coNP}/\mathrm{poly}$~\cite{fomin_lokshtanov_saurabh_zehavi_2019}.

\vspace{4mm}

We also use the \textsc{Edge Multicut} problem as a source problem in our reductions.

\vspace{3mm}

\noindent\fbox{%
  \begin{minipage}{32.4em}
    {\sc Edge Multicut}\\
    \noindent{\bf Input:} A directed graph $G=(V,A)$, a collection 
    $\mathcal{T}\subseteq V\times V$ of ordered terminal pairs, 
    and an integer $k$. \\
    \noindent{\bf Question:} Does there exist a set 
    $F\subseteq A$ of size at most $k$ such that, for every 
    $(s,t)\in\mathcal{T}$, there is no directed path from $s$ to $t$ 
    in $G-F$?
  \end{minipage}%
}

\vspace{2mm}

In the \textsc{Edge Multicut} problem, the goal is to delete at most $k$ arcs 
so as to separate every ordered terminal pair $(s,t)\in\mathcal{T}$, 
that is, to destroy all directed paths from $s$ to $t$. 
The problem is known to be $\mathrm{W[1]}$-hard when parameterized by the size of the cutset $k$, 
even on directed acyclic graphs~\cite{10.1007/978-3-642-31594-7_49}.

\section{W[1]-hardness under Treewidth}

In this section, we prove the following theorem.

\begin{theorem}\label{hardness-result}
{\sc $s$-Club Cluster Edge Deletion} is W[1]-hard when parameterized by the pathwidth of the input graph.
\end{theorem}

\begin{proof}
We prove the theorem by a parameterized reduction from 
\textsc{Minimum Maximum Outdegree}, defined in Section~\ref{problems section}. 
This problem is known to be W[1]-hard when parameterized by the vertex cover number of the input graph~\cite{bodlaender2022problemshardtreewidtheasy}.

We now describe our reduction to 
{\sc $s$-Club Cluster Edge Deletion}.
Recall that $\vc(G)$ and $\pw(G)$ denote the vertex cover number and pathwidth of the graph $G$, respectively.

\proofpara{Constructing the \textsc{$s$-Club Cluster Edge Deletion} instance}
Let the vertex cover number of $G$ be $k$, and let
$C=\{a_1,\ldots,a_k\}$ denote a minimum vertex cover of $G$.
Let $I=\{b_1,\ldots,b_{n'}\}$ be the independent set obtained after removing
the vertex cover $C$, where $n':=n-k$.
Given the input instance $\mathcal{I}=(G,w,r)$, we construct an equivalent instance
$\mathcal{I}'=(G',k',s)$ of {\sc $s$-Club Cluster Edge Deletion}.

\proofpara{Start of the construction}
Let $C=\{a_1,\ldots,a_k\}$ be a minimum vertex cover of graph $G$ and 
$I=\{b_1,\ldots,b_{n'}\}$ be the independent set obtained after removing the vertex cover $C$ where $n':=n-k$. 
We construct a gadget $H_{a_{t}}$ and $H_{b_{s}}$ corresponding to every vertex $a_{t}\in C$ and $b_{s}\in I$ respectively.
Later we describe how these gadgets are connected to each other.

\proofpara{Overview of the construction}
At a high level, the construction creates one gadget for every vertex of the input graph. 
Each gadget contains a long path together with additional vertices encoding the neighborhood structure of the corresponding vertex. 
Whenever two vertices are adjacent in the original graph, the corresponding gadgets are connected by carefully constructed paths. 
The lengths of these paths encode the edge weights of the input instance, while their structure guarantees robustness under a bounded number of edge deletions. 
Overall, the construction is designed so that deleting a small number of edges in the resulting graph corresponds to choosing a valid orientation satisfying the outdegree constraints in the input instance.

\proofpara{Construction of $H_{a_{t}}$}
We describe the construction in three steps.

\smallskip
\noindent\textbf{Step~1 (Spine path construction).}
The gadget contains a path of length $2n+1$. 
The vertices of this path are partitioned into two sets.
The first set is $X^{a_{t}} = \{x^{a_{t}}_{0},x^{a_{t}}_{1},x^{a_{t}}_{2},\ldots,x^{a_{t}}_{n}\}$.
The second set is $Y^{a_{t}}$ which is further partitioned into two sets
$Y^{a_{t}}_{C} = \{ y^{a_{t}}_{a_{1}}, y^{a_{t}}_{a_{2}},\ldots,y^{a_{t}}_{a_{k}}\}$ and 
$Y^{a_{t}}_{I} = \{ y^{a_{t}}_{b_{1}}, y^{a_{t}}_{b_{2}},\ldots,y^{a_{t}}_{b_{n'}}\}$.
The vertices of the sets $X^{a_{t}}$ and $Y^{a_{t}}$ appear alternatively on this path.
Let us describe the adjacency in more detail. 
The vertex $y^{a_{t}}_{a_{i}}$ is adjacent to $x^{a_{t}}_{i-1}$ and $x^{a_{t}}_{i}$ for each $i\in [k]$.
Then the vertex $y^{a_{t}}_{b_{j}}$ is adjacent to $x^{a_{t}}_{k+(j-1)}$ and $x^{a_{t}}_{k+j}$ for each $j\in[n']$.
Please refer to Figure \ref{fig:path-at} for illustration of this path.

\begin{figure}[ht]
\centering
\begin{tikzpicture}[
  x=0.9cm,y=1.2cm,
  vtx/.style={circle,fill=black,inner sep=1pt},
  edg/.style={line width=0.8pt},
  lbx/.style={font=\scriptsize, yshift=2.5mm},
  lby/.style={font=\scriptsize, yshift=-2.5mm}
]

\node[vtx] (x0) at (0,0) {}; \node[lbx] at (x0) {$x^{a_t}_{0}$};
\node[vtx] (y1) at (0.6,-1) {}; \node[lby] at (y1) {$y^{a_t}_{a_{1}}$};
\node[vtx] (x1) at (1.2,0) {}; \node[lbx] at (x1) {$x^{a_t}_{1}$};
\node[vtx] (y2) at (1.8,-1) {}; \node[lby] at (y2) {$y^{a_t}_{a_{2}}$};
\node[vtx] (x2) at (2.4,0) {}; \node[lbx] at (x2) {$x^{a_t}_{2}$};
\node[vtx] (y3) at (3.0,-1) {}; \node[lby] at (y3) {$y^{a_t}_{a_{3}}$};
\node[vtx] (x3) at (3.6,0) {}; \node[lbx] at (x3) {$x^{a_t}_{3}$};
\draw[edg] (x0)--(y1)--(x1)--(y2)--(x2)--(y3)--(x3);

\node at (4.4,-0.5) {$\cdots$};

\node[vtx] (xvc) at (5.2,0) {}; \node[lbx] at (xvc) {$x^{a_t}_{k}$};
\node[vtx] (yvc) at (5.8,-1) {}; \node[lby] at (yvc) {$y^{a_t}_{a_{k}}$};
\node[vtx] (xvc1) at (6.4,0) {}; \node[lbx] at (xvc1) {$x^{a_t}_{k+1}$};
\node[vtx] (yb1) at (7.0,-1) {}; \node[lby] at (yb1) {$y^{a_t}_{b_{1}}$};
\node[vtx] (xvc2) at (7.6,0) {}; \node[lbx] at (xvc2) {$x^{a_t}_{k+2}$};
\draw[edg] (xvc)--(yvc)--(xvc1)--(yb1)--(xvc2);

\node at (8.4,-0.5) {$\cdots$};

\node[vtx] (xnm1) at (9.2,0) {}; \node[lbx] at (xnm1) {$x^{a_t}_{n-2}$};
\node[vtx] (ynm1) at (9.8,-1) {}; \node[lby] at (ynm1) {$y^{a_t}_{b_{n'-1}}$};
\node[vtx] (xn) at (10.4,0) {}; \node[lbx] at (xn) {$x^{a_t}_{n-1}$};
\node[vtx] (yn) at (11.0,-1) {}; \node[lby] at (yn) {$y^{a_t}_{b_{n'}}$};
\node[vtx] (xn0) at (11.6,0) {}; \node[lbx] at (xn0) {$x^{a_t}_{n}$};
\draw[edg] (xnm1)--(ynm1)--(xn)--(yn)--(xn0);

\end{tikzpicture}
\caption{The alternating path in $H_{a_t}$: vertices $X^{a_t}=\{x^{a_t}_0,\ldots,x^{a_t}_n\}$ (top) and $Y^{a_t}=\{y^{a_t}_{a_1},\ldots,y^{a_t}_{a_{k}},y^{a_t}_{b_1},\ldots,y^{a_t}_{b_{n'}}\}$ (bottom) appear alternately.}
\label{fig:path-at}
\end{figure}
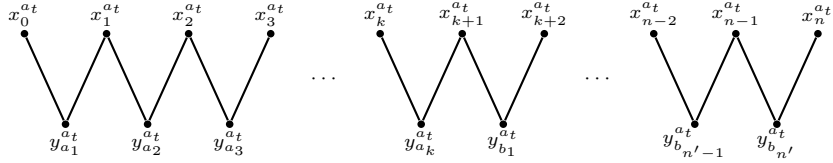

\smallskip
\noindent\textbf{Step~2 (Insertion of reinforced paths).}
Recall from Step~1 that every vertex of $Y^{a_t}$ is adjacent to exactly two
consecutive vertices of the spine $X^{a_t}$. 
For each $i \in [k]$, the vertex $y^{a_t}_{a_i}$ is adjacent to
$x^{a_t}_{i-1}$ and $x^{a_t}_{i}$. 
If $G$ contains the edge $e=\{a_t,a_i\}$, we insert a path of length
$\alpha \cdot wt(e)$ between $x^{a_t}_{i-1}$ and $x^{a_t}_{i}$, where $\alpha$
is an even scaling factor depending only on $r$ and $n$. 
If $e \notin E(G)$, we insert instead a path of length two. 
The construction is analogous for neighbors in the independent set.
For each $j \in [n']$, the vertex $y^{a_t}_{b_j}$ is adjacent to
$x^{a_t}_{k+j-1}$ and $x^{a_t}_{k+j}$; if
$e=\{a_t,b_j\}\in E(G)$ we insert a path of length
$\alpha \cdot wt(e)$, otherwise we insert one of length two.
Finally, we add an auxiliary vertex $x_{0}^{*a_t}$ and connect it to
$x_{0}^{a_t}$ with a path of length $s - (\alpha r+2n+2)$.

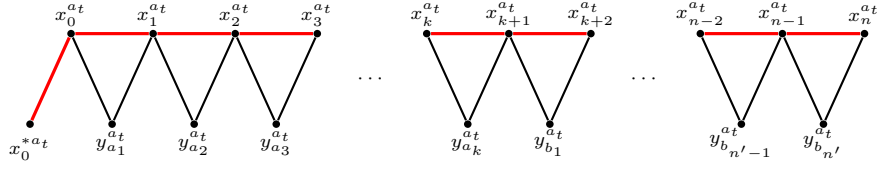
\begin{figure}[ht]
\centering
\begin{tikzpicture}[
  x=0.9cm,y=1.2cm,
  vtx/.style={circle,fill=black,inner sep=1pt},
  edg/.style={line width=0.8pt},
  rededge/.style={very thick,red},
  lbx/.style={font=\scriptsize, yshift=2.5mm},
  lby/.style={font=\scriptsize, yshift=-2.5mm}
]

\node[vtx] (x0) at (0,0) {}; \node[lbx] at (x0) {$x^{a_t}_{0}$};
\node[vtx] (y1) at (0.6,-1) {}; \node[lby] at (y1) {$y^{a_t}_{a_{1}}$};
\node[vtx] (x1) at (1.2,0) {}; \node[lbx] at (x1) {$x^{a_t}_{1}$};
\node[vtx] (y2) at (1.8,-1) {}; \node[lby] at (y2) {$y^{a_t}_{a_{2}}$};
\node[vtx] (x2) at (2.4,0) {}; \node[lbx] at (x2) {$x^{a_t}_{2}$};
\node[vtx] (y3) at (3.0,-1) {}; \node[lby] at (y3) {$y^{a_t}_{a_{3}}$};
\node[vtx] (x3) at (3.6,0) {}; \node[lbx] at (x3) {$x^{a_t}_{3}$};
\node[vtx,label=below:{$x^{*a_t}_0$}] (x0at) at (-0.6,-1) {};
\draw[edg] (x0)--(y1)--(x1)--(y2)--(x2)--(y3)--(x3);

\node at (4.4,-0.5) {$\cdots$};

\node[vtx] (xvc) at (5.2,0) {}; \node[lbx] at (xvc) {$x^{a_t}_{k}$};
\node[vtx] (yvc) at (5.8,-1) {}; \node[lby] at (yvc) {$y^{a_t}_{a_{k}}$};
\node[vtx] (xvc1) at (6.4,0) {}; \node[lbx] at (xvc1) {$x^{a_t}_{k+1}$};
\node[vtx] (yb1) at (7.0,-1) {}; \node[lby] at (yb1) {$y^{a_t}_{b_{1}}$};
\node[vtx] (xvc2) at (7.6,0) {}; \node[lbx] at (xvc2) {$x^{a_t}_{k+2}$};
\draw[edg] (xvc)--(yvc)--(xvc1)--(yb1)--(xvc2);

\node at (8.4,-0.5) {$\cdots$};

\node[vtx] (xnm1) at (9.2,0) {}; \node[lbx] at (xnm1) {$x^{a_t}_{n-2}$};
\node[vtx] (ynm1) at (9.8,-1) {}; \node[lby] at (ynm1) {$y^{a_t}_{b_{n'-1}}$};
\node[vtx] (xn) at (10.4,0) {}; \node[lbx] at (xn) {$x^{a_t}_{n-1}$};
\node[vtx] (yn) at (11.0,-1) {}; \node[lby] at (yn) {$y^{a_t}_{b_{n'}}$};
\node[vtx] (xn0) at (11.6,0) {}; \node[lbx] at (xn0) {$x^{a_t}_{n}$};
\draw[edg] (xnm1)--(ynm1)--(xn)--(yn)--(xn0);

\draw[rededge] (x0) -- (x0at);
\draw[rededge] (x0) -- (x1);
\draw[rededge] (x1) -- (x2);
\draw[rededge] (x3) -- (x2);
\draw[rededge] (xvc) -- (xvc1);
\draw[rededge] (xvc1) -- (xvc2);
\draw[rededge] (xnm1) -- (xn);
\draw[rededge] (xn0) -- (xn);

\end{tikzpicture}
\caption{Step~2 of gadget $H_{a_t}$.  
Starting from the alternating path defined in Step~1, we add reinforced paths (shown in red)  
between consecutive $X^{a_t}$-vertices, and an auxiliary reinforced path connecting $x^{a_t}_0$ to $x^{*a_t}_0$.}
\label{fig:path-at1}
\end{figure}

\smallskip

\noindent\textbf{Step~3 (Augmentation of reinforced paths).}
For every reinforced path introduced in Step~2, we now apply an additional augmentation.  
Let $(v_{0},v_{1},\ldots,v_{\ell})$, where $\ell\ge 2$, denote such a reinforced path.
For each even index $i \in \{0,2,4,\ldots,\ell-2\}$, we insert $k'+1$ internally vertex-disjoint paths of length two between $v_i$ and $v_{i+2}$.  
The purpose of this operation is to guarantee robustness against edge deletions: even if up to $k'$ edges are removed from the graph, the endpoints of every reinforced path remain connected in the resulting graph.  
The precise definition of the parameter $k'$ will be specified later. Figure~\ref{fig:reinforcement-triangles} illustrates these attachments.

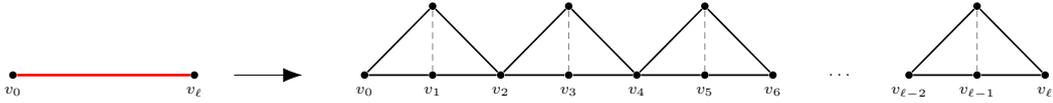
\begin{figure}[ht]
    \centering
\resizebox{\linewidth}{!}{%
\begin{tikzpicture}[
  x=0.9cm,y=1.05cm,
  vtx/.style={circle,fill=black,inner sep=1.2pt},
  base/.style={line width=0.8pt},
  redpath/.style={very thick,red},
  tri/.style={line width=0.75pt},
  mid/.style={densely dashed,gray},
  lab/.style={font=\scriptsize}
]

\node[vtx] (L0) at (0,0) {};
\node[vtx] (Lell) at (3.2,0) {};
\draw[redpath] (L0) -- (Lell);
\node[lab,below=2pt] at (L0) {$v_0$};
\node[lab,below=2pt] at (Lell) {$v_\ell$};

\draw[-{Latex[length=3mm]}] (3.9,0) -- (5.1,0);

\node[vtx] (r0) at (6.2,0)  {}; \node[lab,below=2pt] at (r0) {$v_0$};
\node[vtx] (r1) at (7.4,0)  {}; \node[lab,below=2pt] at (r1) {$v_1$};
\node[vtx] (r2) at (8.6,0)  {}; \node[lab,below=2pt] at (r2) {$v_2$};
\node[vtx] (r3) at (9.8,0)  {}; \node[lab,below=2pt] at (r3) {$v_3$};
\node[vtx] (r4) at (11.0,0) {}; \node[lab,below=2pt] at (r4) {$v_4$};
\node[vtx] (r5) at (12.2,0) {}; \node[lab,below=2pt] at (r5) {$v_5$};
\node[vtx] (r6) at (13.4,0) {}; \node[lab,below=2pt] at (r6) {$v_6$};

\node[lab] (dots) at (14.6,0) {$\cdots$};

\node[vtx] (rl2) at (15.8,0) {}; \node[lab,below=2pt] at (rl2) {$v_{\ell-2}$};
\node[vtx] (rl1) at (17.0,0) {}; \node[lab,below=2pt] at (rl1) {$v_{\ell-1}$};
\node[vtx] (rl)  at (18.2,0) {}; \node[lab,below=2pt] at (rl)  {$v_{\ell}$};

\draw[base] (r0)--(r1)--(r2)--(r3)--(r4)--(r5)--(r6);
\draw[base] (rl2)--(rl1)--(rl);

\node[vtx] (t02) at (7.4,1.05) {};
\draw[tri] (r0)--(t02)--(r2);
\draw[mid] (t02)--(r1);

\node[vtx] (t24) at (9.8,1.05) {};
\draw[tri] (r2)--(t24)--(r4);
\draw[mid] (t24)--(r3);

\node[vtx] (t46) at (12.2,1.05) {};
\draw[tri] (r4)--(t46)--(r6);
\draw[mid] (t46)--(r5);

\node[vtx] (tl2l) at (17.0,1.05) {};
\draw[tri] (rl2)--(tl2l)--(rl);
\draw[mid] (tl2l)--(rl1);

\end{tikzpicture}}
\caption{Step~3. Reinforcement of a path by inserting $k'+1$ disjoint 
2-length connections between every second pair of vertices.}
\label{fig:reinforcement-triangles}
\end{figure}

This finishes the construction of the gadget $H_{a_{t}}$.

\proofpara{Construction of $H_{b_{s}}$}
For every vertex $b_{s} \in I$ we construct a gadget $H_{b_{s}}$ where $s\in[n']$.
The role of this gadget is to encode the adjacency pattern of $b_{s}$ in the original graph $G$. 
As there are no edges within $I$, the gadget $H_{b_s}$ is defined slightly differently.
Thus $H_{b_s}$ contains a path containing the vertices from the set 
$X^{b_s} = \{x^{b_s}_0,x^{b_s}_1,\ldots,x^{b_s}_{\vc(G)}\}$
and
$Y^{b_s} = \{y^{b_s}_{a_1},\ldots,y^{b_s}_{a_{\vc(G)}}\}$.
Each $y^{b_s}_{a_i}$ is adjacent to $x^{b_s}_{i}$ and $x^{b_s}_{i-1}$, and if $G$ contains edge $e=\{b_s,a_i\}$ we insert a {\it reinforced path} of length $\alpha\cdot wt(e)$, 
otherwise we add a {\it reinforced path} of length two.
Finally, add an auxiliary vertex \(x_{0}^{*b_s}\) and connect it to \(x_{0}^{b_s}\) with a {\it reinforced paths} of length \(s-(\alpha r+2n+2)\).
These {\it reinforced paths} are again strengthened as in Step~3.

\smallskip
Having described the construction of all vertex gadgets, we now define how these gadgets are connected according to the adjacencies of the original graph~$G$ to obtain the final graph~$G'$. Figure~\ref{mainfig} provides an overview of the complete construction.

\proofpara{Construction of $G'$}
The graph $G'$ is obtained by assembling all gadgets $\{H_{a_t}\}_{t=1}^{k}$ 
and $\{H_{b_s}\}_{s=1}^{n'}$ described above, together with inter-gadget 
reinforced paths corresponding to the edges of $G$. 

\smallskip
\noindent
\emph{Inter-gadget connections.}  
For every edge $\{u,v\}\in E(G)$, let $u$ be an endpoint belonging to the vertex cover $C$. Recall that gadget $H_u$ contains the vertex $y^{u}_{v}$ adjacent to the spine $X^{u}$, 
and gadget $H_v$ contains the vertex $y^{v}_{u}$ adjacent to the spine $X^{v}$. 
In $G'$ we connect $y^{u}_{v}$ and $y^{v}_{u}$ by a reinforced path 
of length two. Concretely, we add a path of length two between $y^{u}_{v}$ and 
$y^{v}_{u}$ and then apply the reinforcement from Step~3: between every second 
pair of consecutive vertices on this path we insert $k'+1$ internally vertex-disjoint 
paths of length two.
As a result, the vertices $y^{u}_{v}$ and $y^{v}_{u}$ remain 
connected under the deletion of at most $k'$ edges. 

\smallskip
\noindent
\emph{Parameter settings.}  
We set the scaling factor $\alpha := 2(nr)^2$, the solution size 
$k' := 2|E(G)|$, and the diameter bound $s := 100(nr)^4$.

\begin{figure}[ht]
\centering
\begin{tikzpicture}[scale=0.8,
  x=1cm,y=1cm,
  v/.style={circle,draw,fill=white,inner sep=0pt,minimum size=3.2pt},
  ed/.style={line width=.8pt},
  reded/.style={very thick,red},
  lab/.style={font=\scriptsize}
]

\node[v] (a1) at (6,4) {};
\node[v] (a2) at (8,4) {};
\node[lab,above=2pt] at (a1) {$a_{1}$};
\node[lab,above=2pt] at (a2) {$a_{2}$};

\node[v] (b1) at (5,3) {};
\node[v] (b2) at (6,3) {};
\node[v] (b3) at (7,3) {};
\node[v] (b4) at (8,3) {};
\node[lab,left=3pt]  at (b1) {$b_{1}$};
\node[lab,below=2pt] at (b2) {$b_{2}$};
\node[lab,below=2pt] at (b3) {$b_{3}$};
\node[lab,right=3pt] at (b4) {$b_{4}$};

\draw[ed] (a1)--(a2);
\draw[ed] (a1)--(b1);
\draw[ed] (a1)--(b2);
\draw[ed] (b2)--(a2);
\draw[ed] (b3)--(a2);
\draw[ed] (b4)--(a2);
\draw[ed] (a1)--(b3);

\node[lab] at (7,2) {(a) Graph $G$};

\node[v] (xa10) at (0,0) {};
\node[v] (xa11) at (1,0) {};
\node[v] (xa12) at (2,0) {};
\node[v] (xa13) at (3,0) {};
\node[v] (xa14) at (4,0) {};
\node[v] (xa15) at (5,0) {};
\node[v] (xa16) at (6,0) {};
\node[lab,above=2pt] at (xa10) {$x_{0}^{a_{1}}$};
\node[lab,above=2pt] at (xa11) {$x_{1}^{a_{1}}$};
\node[lab,above=2pt] at (xa12) {$x_{2}^{a_{1}}$};
\node[lab,above=2pt] at (xa13) {$x_{3}^{a_{1}}$};
\node[lab,above=2pt] at (xa14) {$x_{4}^{a_{1}}$};
\node[lab,above=2pt] at (xa15) {$x_{5}^{a_{1}}$};
\node[lab,above=2pt] at (xa16) {$x_{6}^{a_{1}}$};

\node[v] (x01)    at (-.5,-1) {};
\node[v] (ya1a1)  at ( .5,-1) {};
\node[v] (ya1a2)  at (1.5,-1) {};
\node[v] (ya1b1)  at (2.5,-1) {};
\node[v] (ya1b2)  at (3.5,-1) {};
\node[v] (ya1b3)  at (4.5,-1) {};
\node[v] (ya1b4)  at (5.5,-1) {};
\node[lab,left=3pt]  at (x01)   {$x_{0}^{* a_{1}}$};
\node[lab,right=3pt] at (ya1a1) {$y_{a_1}^{a_{1}}$};
\node[lab,right=3pt] at (ya1a2) {$y_{a_2}^{a_{1}}$};
\node[lab,right=3pt] at (ya1b1) {$y_{b_1}^{a_{1}}$};
\node[lab,right=3pt] at (ya1b2) {$y_{b_2}^{a_{1}}$};
\node[lab,right=3pt] at (ya1b3) {$y_{b_3}^{a_{1}}$};
\node[lab,right=3pt] at (ya1b4) {$y_{b_{4}}^{a_{1}}$};

\draw[reded] (xa10)--(x01);
\draw[reded] (xa10)--(xa11)--(xa12)--(xa13)--(xa14)--(xa15)--(xa16);

\draw[ed] (xa10)--(ya1a1);
\draw[ed] (xa11)--(ya1a1);
\draw[ed] (xa11)--(ya1a2);
\draw[ed] (xa12)--(ya1a2);
\draw[ed] (xa12)--(ya1b1);
\draw[ed] (xa13)--(ya1b1);
\draw[ed] (xa13)--(ya1b2);
\draw[ed] (xa14)--(ya1b2);
\draw[ed] (xa14)--(ya1b3);
\draw[ed] (xa15)--(ya1b3);
\draw[ed] (xa15)--(ya1b4);
\draw[ed] (xa16)--(ya1b4);

\begin{scope}[on background layer]
  \draw[
    rounded corners=3pt,
    draw=blue!60,
    fill=blue!20,
    fill opacity=0.18,
    line width=0.6pt
  ] (-1.8,-1.5) rectangle (6.35,0.8);
  \node[anchor=south west,font=\scriptsize,blue!60] at (2.5,0.8) {$H_{a_1}$};
\end{scope}

\node[v] (xa20) at (8,0) {};
\node[v] (xa21) at (9,0) {};
\node[v] (xa22) at (10,0) {};
\node[v] (xa23) at (11,0) {};
\node[v] (xa24) at (12,0) {};
\node[v] (xa25) at (13,0) {};
\node[v] (xa26) at (14,0) {};
\node[lab,above=2pt] at (xa20) {$x_{0}^{a_{2}}$};
\node[lab,above=2pt] at (xa21) {$x_{1}^{a_{2}}$};
\node[lab,above=2pt] at (xa22) {$x_{2}^{a_{2}}$};
\node[lab,above=2pt] at (xa23) {$x_{3}^{a_{2}}$};
\node[lab,above=2pt] at (xa24) {$x_{4}^{a_{2}}$};
\node[lab,above=2pt] at (xa25) {$x_{5}^{a_{2}}$};
\node[lab,above=2pt] at (xa26) {$x_{6}^{a_{2}}$};

\node[v] (x02)    at (7.7,-1) {};
\node[v] (ya2a1)  at (8.5,-1) {};
\node[v] (ya2a2)  at (9.5,-1) {};
\node[v] (ya2b1)  at (10.5,-1) {};
\node[v] (ya2b2)  at (11.5,-1) {};
\node[v] (ya2b3)  at (12.5,-1) {};
\node[v] (ya2b4)  at (13.5,-1) {};
\node[lab,left=3pt]  at (x02)   {$x_{0}^{* a_{2}}$};
\node[lab,right=3pt] at (ya2a1) {$y_{a_1}^{a_{2}}$};
\node[lab,right=3pt] at (ya2a2) {$y_{a_2}^{a_{2}}$};
\node[lab,right=3pt] at (ya2b1) {$y_{b_1}^{a_{2}}$};
\node[lab,right=3pt] at (ya2b2) {$y_{b_2}^{a_{2}}$};
\node[lab,right=3pt] at (ya2b3) {$y_{b_3}^{a_{2}}$};
\node[lab,right=3pt] at (ya2b4) {$y_{b_{4}}^{a_{2}}$};

\draw[reded] (xa20)--(x02);
\draw[reded] (xa20)--(xa21)--(xa22)--(xa23)--(xa24)--(xa25)--(xa26);

\draw[ed] (xa20)--(ya2a1);
\draw[ed] (xa21)--(ya2a1);
\draw[ed] (xa21)--(ya2a2);
\draw[ed] (xa22)--(ya2a2);
\draw[ed] (xa22)--(ya2b1);
\draw[ed] (xa23)--(ya2b1);
\draw[ed] (xa23)--(ya2b2);
\draw[ed] (xa24)--(ya2b2);
\draw[ed] (xa24)--(ya2b3);
\draw[ed] (xa25)--(ya2b3);
\draw[ed] (xa25)--(ya2b4);
\draw[ed] (xa26)--(ya2b4);

\begin{scope}[on background layer]
  \draw[
    rounded corners=3pt,
    draw=blue!60,
    fill=blue!20,
    fill opacity=0.18,
    line width=0.6pt
  ] (6.7,-1.5) rectangle (14.35,0.8);
  \node[anchor=south west,font=\scriptsize,blue!60] at (11,0.8) {$H_{a_2}$};
\end{scope}

\node[v] (xb10) at (0,-6) {};
\node[v] (xb11) at (1,-6) {};
\node[v] (xb12) at (2,-6) {};
\node[lab,below=2pt] at (xb10) {$x_{0}^{b_{1}}$};
\node[lab,below=2pt] at (xb11) {$x_{1}^{b_{1}}$};
\node[lab,below=2pt] at (xb12) {$x_{2}^{b_{1}}$};

\node[v] (y0b1)  at (-.5,-5) {};
\node[v] (yb1a1) at ( .5,-5) {};
\node[v] (yb1a2) at (1.5,-5) {};
\node[lab,left=3pt] at (y0b1)  {$x_{0}^{* b_{1}}$};
\node[lab,right=3pt] at (yb1a1) {$y_{a_1}^{b_{1}}$};
\node[lab,right=3pt] at (yb1a2) {$y_{a_2}^{b_{1}}$};

\draw[reded] (y0b1)--(xb10);
\draw[reded] (xb11)--(xb10);
\draw[reded] (xb12)--(xb11);
\draw[ed] (xb10)--(yb1a1);
\draw[ed] (xb11)--(yb1a1);
\draw[ed] (xb11)--(yb1a2);
\draw[ed] (xb12)--(yb1a2);

\begin{scope}[on background layer]
  \draw[
    rounded corners=3pt,
    draw=blue!60,
    fill=blue!20,
    fill opacity=0.18,
    line width=0.6pt
  ] (-1.8,-6.9) rectangle (2.35,-4.5);
  \node[anchor=south west,font=\scriptsize,blue!60] at (0,-4.5) {$H_{b_1}$};
\end{scope}

\node[v] (xb20) at (4,-6) {};
\node[v] (xb21) at (5,-6) {};
\node[v] (xb22) at (6,-6) {};
\node[lab,below=2pt] at (xb20) {$x_{0}^{b_{2}}$};
\node[lab,below=2pt] at (xb21) {$x_{1}^{b_{2}}$};
\node[lab,below=2pt] at (xb22) {$x_{2}^{b_{2}}$};

\node[v] (y0b2)  at (3.2,-5) {};
\node[v] (yb2a1) at (4.5,-5) {};
\node[v] (yb2a2) at (5.5,-5) {};
\node[lab,right=3pt] at (y0b2)  {$x_{0}^{* b_{2}}$};
\node[lab,right=3pt] at (yb2a1) {$y_{a_1}^{b_{2}}$};
\node[lab,right=3pt] at (yb2a2) {$y_{a_2}^{b_{2}}$};

\draw[very thick, red] (y0b2) -- (xb20);
\draw[reded] (xb21)--(xb20);
\draw[reded] (xb22)--(xb21);
\draw[ed] (xb20)--(yb2a1);
\draw[ed] (xb21)--(yb2a1);
\draw[ed] (xb21)--(yb2a2);
\draw[ed] (xb22)--(yb2a2);

\begin{scope}[on background layer]
  \draw[
    rounded corners=3pt,
    draw=blue!60,
    fill=blue!20,
    fill opacity=0.18,
    line width=0.6pt
  ] (3,-6.9) rectangle (6.5,-4.5);
  \node[anchor=south west,font=\scriptsize,blue!60] at (5,-4.5) {$H_{b_2}$};
\end{scope}

\node[v] (xb30) at (8,-6) {};
\node[v] (xb31) at (9,-6) {};
\node[v] (xb32) at (10,-6) {};
\node[lab,below=2pt] at (xb30) {$x_{0}^{b_{3}}$};
\node[lab,below=2pt] at (xb31) {$x_{1}^{b_{3}}$};
\node[lab,below=2pt] at (xb32) {$x_{2}^{b_{3}}$};

\node[v] (y0b3)  at (7.2,-5) {};
\node[v] (yb3a1) at (8.5,-5) {};
\node[v] (yb3a2) at (9.5,-5) {};
\node[lab,right=3pt] at (y0b3)  {$x_{0}^{* b_{3}}$};
\node[lab,right=3pt] at (yb3a1) {$y_{a_1}^{b_{3}}$};
\node[lab,right=3pt] at (yb3a2) {$y_{a_2}^{b_{3}}$};

\draw[reded] (y0b3)--(xb30);
\draw[reded] (xb31)--(xb30);
\draw[reded] (xb32)--(xb31);
\draw[ed] (xb30)--(yb3a1);
\draw[ed] (xb31)--(yb3a1);
\draw[ed] (xb31)--(yb3a2);
\draw[ed] (xb32)--(yb3a2);

\begin{scope}[on background layer]
  \draw[
    rounded corners=3pt,
    draw=blue!60,
    fill=blue!20,
    fill opacity=0.18,
    line width=0.6pt
  ] (7,-6.9) rectangle (10.5,-4.5);
  \node[anchor=south west,font=\scriptsize,blue!60] at (8.8,-4.5) {$H_{b_3}$};
\end{scope}

\node[v] (xb40) at (12,-6) {};
\node[v] (xb41) at (13,-6) {};
\node[v] (xb42) at (14,-6) {};
\node[lab,below=2pt] at (xb40) {$x_{0}^{b_{4}}$};
\node[lab,below=2pt] at (xb41) {$x_{1}^{b_{4}}$};
\node[lab,below=2pt] at (xb42) {$x_{2}^{b_{4}}$};

\node[v] (y0b4)  at (11.2,-5) {};
\node[v] (yb4a1) at (12.5,-5) {};
\node[v] (yb4a2) at (13.5,-5) {};
\node[lab,right=3pt] at (y0b4)  {$x_{0}^{* b_{4}}$};
\node[lab,right=3pt] at (yb4a1) {$y_{a_1}^{b_{4}}$};
\node[lab,right=3pt] at (yb4a2) {$y_{a_2}^{b_{4}}$};

\draw[reded] (y0b4)--(xb40);
\draw[reded] (xb41)--(xb40);
\draw[reded] (xb42)--(xb41);
\draw[ed] (xb40)--(yb4a1);
\draw[ed] (xb41)--(yb4a1);
\draw[ed] (xb41)--(yb4a2);
\draw[ed] (xb42)--(yb4a2);

\begin{scope}[on background layer]
  \draw[
    rounded corners=3pt,
    draw=blue!60,
    fill=blue!20,
    fill opacity=0.18,
    line width=0.6pt
  ] (11,-6.9) rectangle (14.5,-4.5);
  \node[anchor=south west,font=\scriptsize,blue!60] at (12.6,-4.5) {$H_{b_4}$};
\end{scope}

\draw[reded] (yb1a1)--(ya1b1);
\draw[reded] (yb2a1)--(ya1b2);
\draw[reded] (yb2a2)--(ya2b2);
\draw[reded] (yb3a2)--(ya2b3);
\draw[reded] (yb4a2)--(ya2b4);
\draw[reded] (yb3a1)--(ya1b3);
\draw[reded] (ya2a1) ..controls(5,-3).. (ya1a2);

\node[lab] at (7,-7.3) {(b) Graph $G'$};

\end{tikzpicture}
\caption{Illustration of the reduction. 
(a) Input graph $G$ on vertex cover $\{a_1,a_2\}$ and independent set $\{b_1,\ldots,b_4\}$.  
(b) Constructed graph $G'$, showing the gadgets $H_{a_t}$ (middle), $H_{b_s}$ (bottom), and 
the reinforced paths (in red) introduced both within gadgets and between them. 
For readability, each reinforced path is drawn as a single thick red edge, 
though in the construction it is expanded as described in Steps~2--3.} 
\label{mainfig}
\end{figure}
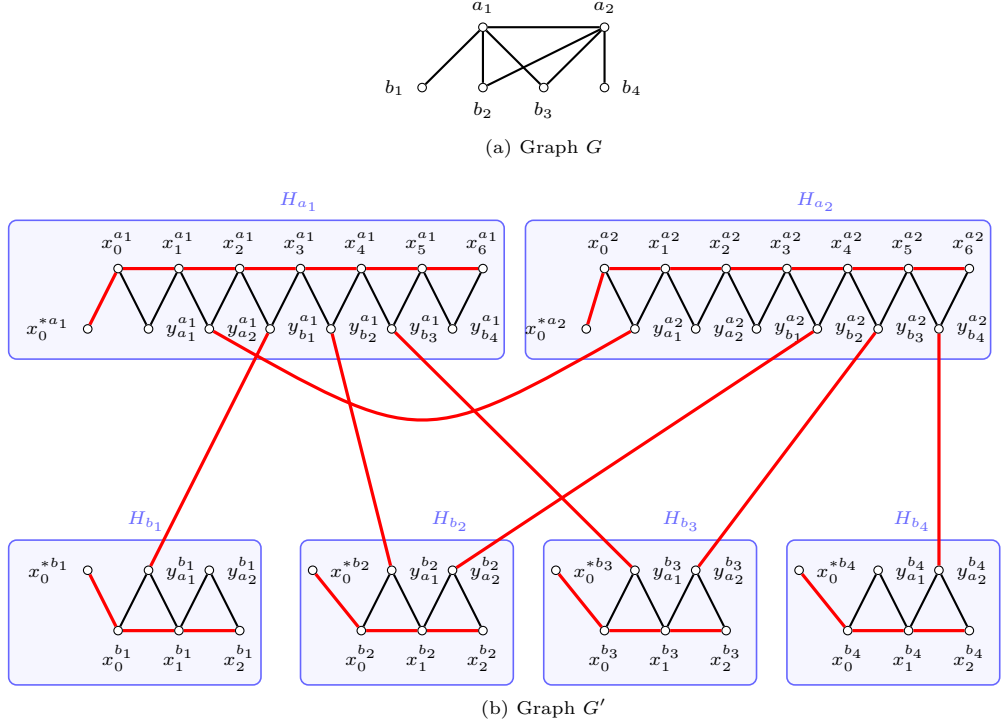

\proofpara{Pathwidth of $G'$}
We now argue that the constructed graph $G'$ has pathwidth bounded by $O(k^2)$, 
where $k=\vc(G)$. The proof proceeds in three steps.

\smallskip
\noindent\textbf{Step~1: Compression to $G''$.}
Obtain $G''$ from $G'$ by replacing every reinforced path (of any length,
within a gadget or between two gadgets) by a single edge between its
endpoints. We claim that $G''$ admits a path decomposition whose bags
have size at most $3k^2+k$.
Recall that $C=\{a_1,\dots,a_k\}$ is a vertex cover and $I=\{b_1,\dots,b_{n'}\}$ is the independent set in $G$. 

\emph{Bag $B_0$.}  
For each $t\in[k]$, include all vertices of $Y^{a_t}_C$, the two neighbors of
each $y^{a_t}_{a_i}$ in $X^{a_t}$, and the auxiliary vertex $x^{*a_t}_0$.  
Altogether,
$B_0 \;=\; \bigcup_{t=1}^k \Big( Y^{a_t}_C \cup N_{X^{a_t}}(Y^{a_t}_C) \cup \{x^{*a_t}_0\} \Big)$,
where $N_{X^{a_t}}(Y^{a_t}_C)$ denotes the two unique $X^{a_t}$-neighbors of
each $y^{a_t}_{a_i}$.  
Since $|Y^{a_t}_C|=k$, these contribute $k$ vertices per $t$. 
Each $y^{a_t}_{a_i}$ has two neighbors in $X^{a_t}$, so in total at most $2k$ further vertices appear in $N_{X^{a_t}}(Y^{a_t}_C)$; together with the auxiliary $x^{*a_t}_0$ this gives at most $k+2k+1=3k+1$ vertices from each gadget. 
Summing over all $t\in[k]$ yields an upper bound $|B_0|\le 3k^2+k$.

\smallskip
\emph{Bags $B_j$.}  
For each $j\in[n']$, include the vertices $\{y^{a_t}_{b_j}:t\in[k]\}$,
their two unique neighbors in $X^{a_t}$, and all vertices of the gadget $H_{b_j}$
in $G''$. 
Thus $B_j \;=\; \{y^{a_t}_{b_j}:t\in[k]\}\;\cup\;N_X(\{y^{a_t}_{b_j}:t\in[k]\})\;\cup\;V(H_{b_j})$,
and since $|V(H_{b_j})|\le 2k+1$, it follows that $|B_j|\le 3k+(2k+1) \le 5k+1$.

\emph{Ordering and contiguity.}
Consider the bag sequence $B_0,\;B_1,\;B_2,\;\ldots,\;B_{n'}$.
We verify the two path–decomposition conditions.

\smallskip
\noindent\textbf{(P1) Edge coverage.}
All edges of $G''$ fall into the following types, each covered by at least one bag:
\begin{itemize}
  \item \emph{Edges inside $H_{b_j}$} (in $G''$): both endpoints lie in $V(H_{b_j})\subseteq B_j$.
  \item \emph{Compressed inter-gadget edges} $\{y^{a_t}_{b_j},y^{b_j}_{a_t}\}$: by definition $y^{a_t}_{b_j}\in B_j$ and $y^{b_j}_{a_t}\in V(H_{b_j})\subseteq B_j$.
  \item \emph{Attachment edges on the $I$-side} 
        $\{y^{a_t}_{b_j},x^{a_t}_{k+j-1}\}$ and $\{y^{a_t}_{b_j},x^{a_t}_{k+j}\}$:
        all three vertices are in $B_j$.
  \item \emph{Attachment edges on the $C$-side} 
        $\{y^{a_t}_{a_i},x^{a_t}_{i-1}\}$ and $\{y^{a_t}_{a_i},x^{a_t}_{i}\}$:
        all three vertices are in $B_0$.
  \item \emph{Compressed spine edges within $X^{a_t}$}:  
        for $1\le i\le k$ the edge $\{x^{a_t}_{i-1},x^{a_t}_{i}\}$ is covered by $B_0$ (both endpoints appear there as neighbors of $y^{a_t}_{a_i}$);  
        for $k<i\le k+n'$ write $i=k+j$ and the edge $\{x^{a_t}_{i-1},x^{a_t}_{i}\}=\{x^{a_t}_{k+j-1},x^{a_t}_{k+j}\}$ is covered by $B_j$.
  \item \emph{Auxiliary edge} $\{x^{*a_t}_0,x^{a_t}_0\}$: both vertices are in $B_0$.
\end{itemize}
Thus every edge of $G''$ is contained in some bag.

\smallskip
\noindent\textbf{(P2) Vertex contiguity.}
For each vertex $v$, the set of indices of bags containing $v$ forms an interval:
\begin{itemize}
  \item If $v\in V(H_{b_j})$, then $v$ appears only in $B_j$.
  \item If $v=y^{a_t}_{b_j}$, then $v$ appears only in $B_j$.
  \item If $v=y^{a_t}_{a_i}$ or $v=x^{*a_t}_0$, then $v$ appears only in $B_0$.
  \item If $v=x^{a_t}_i\in X^{a_t}$, then:
    \begin{itemize}
      \item for $0\le i\le k$, $v$ is in $B_0$ (as a neighbor of some $y^{a_t}_{a_i}$);
            among the $I$-side attachments, $x^{a_t}_k$ also appears in $B_1$, so
            $v$ occurs in either $\{0\}$ or $\{0,1\}$;
      \item for $i>k$, write $i=k+j$; then $v$ appears in $B_j$ (as a neighbor of $y^{a_t}_{b_j}$) and,
            if $j<n'$, also in $B_{j+1}$ (as a neighbor of $y^{a_t}_{b_{j+1}}$). Hence $v$ occurs in
            $\{j\}$ or $\{j,j+1\}$. 
    \end{itemize}
\end{itemize}
In all cases the indices form a contiguous interval. Therefore the sequence $(B_0,B_1,\ldots,B_{n'})$ is a valid
path decomposition.

\smallskip
\noindent\textbf{Bag size.}
By the preceding bounds we have $\max\{|B_0|,|B_j|:j\in[n']\}\le \max\{\,3k^2+k,\;5k+1\,\}$, so the width is at most this maximum minus $1$.

\smallskip
\noindent\textbf{Step~2: Expansion back to $G'$.}
Each compressed edge $\{a,b\}$ of $G''$ corresponds to a reinforced path
$P=(a=z_0,z_1,\dots,z_\ell=b)$ in $G'$, possibly with $k'+1$ parallel
length-2 shortcuts (Step~3 of the construction).  
Choose a bag containing $\{a,b\}$ and replace it with the sequence
$\{a,z_1,b\},\; \{z_1,z_2,b\},\;\dots,\;$ $\{z_{\ell-1},z_\ell=b\}$,
maintaining one endpoint until the other is reached.  
This increases bag size by at most $+2$.  

\smallskip
\noindent\textbf{Step~3: Conclusion.}
Combining the bounds,
$\pw(G') \;\le\; \pw(G'')+2
   \;\le\; \big(\max\{3k^2+k,\;5k+1\}-1\big)+2
   \;=\; O(k^2)$.

\proofpara{Equivalence of $I$ and $I'$}  
We now prove that the reduction is correct, i.e.\ the original instance $I$ 
has a solution if and only if the constructed instance $I'$ has a solution.  

\proofpara{Forward direction: constructing $F$ from $\Lambda$}
Assume $I=(G,w,r)$ is a yes-instance, i.e.\ there exists an orientation
$\Lambda$ of $G$ with $w_{\mathrm{out}}(u)\le r$ for every $u\in V(G)$.
For each oriented edge $(u,v)\in\Lambda$ consider the attachment vertex
$y^u_v$ in the gadget $H_u$. Recall that $y^u_v$ has exactly two
neighbors in the spine $X^u$, which we denote by
$N_{X^u}(y^u_v) = \{\, x^u_{\mathrm{L}}(v),\, x^u_{\mathrm{R}}(v) \,\}$.
We define $F \;:=\; \bigcup_{(u,v)\in\Lambda}\{\, \{y^u_v, x\} : x\in N_{X^u}(y^u_v) \,\}$.
That is, for every oriented edge $(u,v)$ we add to $F$ the two
edges joining $y^u_v$ to its spine neighbors in $X^u$.
By construction, for each undirected edge $\{u,v\}\in E(G)$ exactly one
of $(u,v)$ or $(v,u)$ is in $\Lambda$, and hence exactly two edges are
added to $F$.
Therefore $|F| \;=\; 2|E(G)| \;=\; k'$.

\proofpara{Connected components of $G'-F$}
We now describe the components of $G'-F$.  
Fix $u\in V(G)$.  In the gadget $H_u$ we recall:
\begin{itemize}
  \item the spine $X^u=\{x^u_0,\dots,x^u_n\}$,
  \item for $i=1,\dots,n$, the reinforced path $P^u_i$ with endpoints
        $x^u_{i-1},x^u_i$,
  \item the auxiliary reinforced path $P^u_{\mathrm{aux}}$ between
        $x^u_0$ and $x^{*u}_0$, now of length $s-(\alpha r + 2n +2)$,
  \item for each $w\in N_G(u)$, the reinforced inter–gadget path $R^{u,w}$
        between $y^u_w\in H_u$ and $y^w_u\in H_w$.
\end{itemize}

Define
$\mathrm{In}_{\Lambda}(u)=\{w\in N_G(u):(w,u)\in\Lambda\}$, 
and
$\mathrm{Out}_{\Lambda}(u)=\{w\in N_G(u):(u,w)\in\Lambda\}$.
By the definition of $F$, every $y^u_w$ with $(u,w)\in\Lambda$ is detached
from $X^u$, while for $(w,u)\in\Lambda$ the corresponding $y^w_u$ is detached
from $X^w$. Thus the component $C_u$ containing $X^u$ consists of
\begin{align*}
V(C_u)=\;&X^u \;\cup\; V(P^u_{\mathrm{aux}})\;\cup\;\bigcup_{i=1}^n V(P^u_i)
\;\cup\;\{y^u_w : w\in N_G(u)\setminus \mathrm{Out}_{\Lambda}(u)\}\\
&\;\cup\;\bigcup_{w\in \mathrm{In}_{\Lambda}(u)}\bigl(V(R^{u,w})\cup\{y^w_u\}\bigr).
\end{align*}
Moreover, $C_u\neq C_v$ for distinct $u,v$, since every edge $\{u,v\}\in E(G)$
has exactly one side detached from its spine, preventing connectivity.

\proofpara{Bounding the diameter}
We now prove that every connected component $C_u$ of $G'-F$ has diameter at most $s$. 
Fix $u\in V(G)$. Observe that the vertices farthest from $x^{*u}_0$ lie on the spine of the gadget, while all other vertices are attached to the spine by paths of bounded length. Thus, it suffices to bound the distance from $x^{*u}_0$ to the vertices on the spine and its adjacent reinforced paths.

The auxiliary reinforced path between $x^{*u}_0$ and $x^u_0$ has length
$s-(\alpha r+2n+2)$. From $x^u_0$, traversing the entire spine and all reinforced segments contributes at most $\alpha r+2n$ additional edges: at most $\alpha r$ from edge-encoding reinforced paths and at most $2n$ from the length--2 reinforced paths. Hence every vertex
$z\in X^u\cup\bigcup_i V(P^u_i)$ satisfies
\[
\dist(x^{*u}_0,z)
\le
(s-\alpha r-2n-2)+(\alpha r+2n)
\le s.
\]

Finally, every inter-gadget reinforced path $R^{u,w}$ has length two and is attached to a vertex already lying within distance at most
$(s-\alpha r-2n-2)+(\alpha r+2n)$ from $x^{*u}_0$. Therefore every vertex of
$R^{u,w}\cup\{y^u_w,y^w_u\}$ is also at distance at most $s$ from $x^{*u}_0$.
Thus every vertex of $C_u$ lies within distance at most $s$ from $x^{*u}_0$, implying that $\operatorname{diam}(C_u)\le s$. Since this holds for every $u\in V(G)$, each connected component of $G'-F$ has diameter at most $s$.

\proofpara{Reverse direction: extracting an orientation from $F$}
Suppose that $I'=(G',s,k')$ is a yes-instance.  
That is, there exists a set $F\subseteq E(G')$ with $|F|\le 2|E(G)|=k'$ such that every connected component of $G'-F$ has diameter at most $s$.

\smallskip

\proofpara{Overview of the reverse direction}
We now explain how to extract from $F$ a valid orientation $\Lambda$ of the original graph $G$. 
First, we show that the spine vertices of each gadget remain connected and therefore define a unique component $C_u$. 
Next, we prove that distinct gadgets cannot belong to the same component, which forces $F$ to sever exactly one side of every inter-gadget attachment. 
This naturally determines an orientation of each edge of $G$. 
Finally, we show that if some vertex had weighted outdegree larger than $r$, then the corresponding gadget would contain two vertices at distance greater than $s$, contradicting the correctness of $F$.

\proofpara{All vertices of $X^u$ lie in a single component $C_u$}
Fix $u\in V(G)$. By Step~2 of the construction, for every consecutive pair
$x^{u}_{i-1},x^{u}_{i}\in X^{u}$ we inserted a reinforced path whose endpoints
are $x^{u}_{i-1}$ and $x^{u}_{i}$. Each reinforced path consists of $k'+1$
edge-disjoint paths between its endpoints. Consequently, deleting at most $k'$
edges from $G'$ cannot disconnect $x^{u}_{i-1}$ and $x^{u}_{i}$. All its edges
are internal to that path; in particular, the only edges that
touch $X^u$ but are \emph{not} within these internal paths are the two
attachment edges incident with each $y^{u}_{w}$ (one to $x^{u}_{i-1}$
and one to $x^{u}_{i}$ for the appropriate $i$). Therefore, regardless of which
edges of $F$ are deleted, all vertices of $X^u$ remain mutually connected in
$G'-F$. Let $C_u$ denote the (unique) connected component of $G'-F$ that
contains $X^{u}$.

\proofpara{Distinct spines induce distinct components}
Let $u\neq v$. Any edge between gadgets $H_u$ and $H_v$ uses only the
pair of attachment vertices $y^{u}_{v}\in H_u$ and $y^{v}_{u}\in H_v$
and the reinforced path between them. To move from $X^{u}$ into that
inter-gadget path one must traverse one of the two attachment edges
incident with $y^{u}_{v}$ and an endpoint in $X^{u}$; symmetrically for
the $X^{v}$-side. If \emph{both} pairs of attachment edges (those
incident with $y^{u}_{v}$ and those incident with $y^{v}_{u}$) survive
in $G'-F$, then $X^{u}$ and $X^{v}$ become connected through the
surviving reinforced path between $y^{u}_{v}$ and $y^{v}_{u}$. In that
case, the component containing $X^{u}$ and $X^{v}$ would also contain the
auxiliary vertices $x^{*u}_0$ and $x^{*v}_0$, and its diameter in $G'$
would be at least $2\bigl(s-\alpha r-2n-2\bigr) \;>\; s$,
since each of $x^{*u}_0$ and $x^{*v}_0$ is at distance
$s-\alpha r-2n-2$ from $x^{u}_0$ and $x^{v}_0$, respectively, and the
inter-gadget reinforced path between $y^{u}_{v}$ and $y^{v}_{u}$ has
length~$2$. This contradicts the assumption that all components of
$G'-F$ have diameter at most $s$. Hence for every $\{u,v\}\in E(G)$,
\emph{at least one} of the two attachment pairs must be severed in $F$,
and, in particular, $C_u\neq C_v$.

\proofpara{Necessary deletions at each inter-gadget interface}
Fix an edge $\{u,v\}\in E(G)$. From the previous paragraph, $F$ must
delete one full attachment on at least one side to avoid merging $C_u$
and $C_v$. Concretely, $F$ must contain either the two unique edges
incident with $y^{u}_{v}$ whose other endpoints lie in $X^{u}$ \emph{or}
the two unique edges incident with $y^{v}_{u}$ whose other endpoints lie
in $X^{v}$. If neither pair were deleted, the reinforced path between
$y^{u}_{v}$ and $y^{v}_{u}$ would survive and would connect $C_u$ and
$C_v$, yielding a component of diameter $>s$ as argued above. Since
$|F|\le 2|E(G)|$, exactly one such pair is deleted for each $\{u,v\}$.
This necessary condition is the basis for extracting an orientation
$\Lambda$ in the next step.

\proofpara{Defining the orientation $\Lambda$ from $F$}
We now construct an orientation $\Lambda$ of the original graph $G$ from the
structure of $F$. Consider any edge $\{u,v\}\in E(G)$. By the argument above,
the deletion set $F$ must contain \emph{exactly one} of the two possible
attachment pairs:
\begin{itemize}
  \item either both edges $\{y^u_v, x\}$ with $x\in N_{X^u}(y^u_v)$, or
  \item both edges $\{y^v_u, x\}$ with $x\in N_{X^v}(y^v_u)$.
\end{itemize}
In the first case, $y^u_v$ is detached from $X^u$, while $y^v_u$ remains
attached to $X^v$; we then orient the edge of $G$ as $(u,v)\in\Lambda$.  
In the second case, $y^v_u$ is detached from $X^v$ while $y^u_v$ remains
attached to $X^u$; we orient the edge of $G$ as $(v,u)\in\Lambda$.
\smallskip

\noindent Thus every undirected edge $\{u,v\}\in E(G)$ receives a unique orientation in
$\Lambda$. Since $|F|\le 2|E(G)|$ and exactly two edges are consumed per
undirected edge, this orientation is well defined for all of $E(G)$.

\proofpara{Bounding the outdegree in $\Lambda$}
It remains to show that the orientation $\Lambda$ constructed above satisfies
$w_{\mathrm{out}}(u)\le r$ for every $u\in V(G)$.
Fix $u\in V(G)$ and consider its component $C_u$ in $G'-F$.  Recall that
the auxiliary reinforced path $P^u_{\mathrm{aux}}$ from $x^{*u}_0$ to $x^u_0$
has length $s-(\alpha r + 2n + 2)$, every non–outgoing segment between
consecutive vertices of $X^u$ can be traversed in at most 2 steps (either
because no edge exists or via $y^u_w$ when $(w,u)\in\Lambda$), and every
outgoing edge \((u,v)\in\Lambda\) contributes a segment of length
$\alpha\cdot wt(u,v)$ along the spine.

Suppose, for a contradiction, that $w_{\mathrm{out}}(u)>r$. As edge weights are
integers, $w_{\mathrm{out}}(u)\ge r+1$, so the total length contributed by
outgoing segments along the spine is at least $\alpha(r+1)$. Therefore the
distance from $x^{*u}_0$ to the endpoint $x^u_n$ of the spine satisfies
\[
\begin{aligned}
\operatorname{dist}_{G'-F}(x^{*u}_0, x^u_n)
&\ge
\underbrace{s-(\alpha r + 2n + 2)}_{\text{aux path}}
+
\underbrace{\alpha(r+1)}_{\text{outgoing segments}} \\
&= s + \alpha - (2n + 2)\\
&> s.
\end{aligned}
\]
since $\alpha$ is an even scaling factor chosen sufficiently large. This
contradicts the assumption that every component of $G'-F$ has diameter at most
$s$. Hence $w_{\mathrm{out}}(u)\le r$ for all $u$, as required.
This finishes the proof of Theorem \ref{hardness-result}.
\end{proof}

To further understand the structural limits of tractability, we also consider a natural 
refinement of the problem where, in addition to bounding the diameter of each component, 
we also restrict their total number. 
This parameter, denoted by~$d$, measures the maximum number of $s$-clubs allowed 
in the resulting graph after at most $k$ edge deletions.  
Intuitively, the parameter~$d$ captures a form of global clustering constraint:
while the classical version asks only for every component to be an $s$-club, 
the $d$-bounded variant requires that all such $s$-clubs together form 
a limited number of cohesive groups.
This variant naturally generalizes the notion of distance-based clustering 
and provides a finer control over the decomposition structure of the solution.

\vspace{3mm}
\noindent\fbox{%
  \begin{minipage}{32.4em}\label{s-club-d-cluster}
    {\sc $s$-Club $d$-Cluster Edge Deletion}\\[1mm]
    \noindent{\bf Input:} A graph $G = (V,E)$, and integers $k, s, d \ge 1$.\\
    \noindent{\bf Question:} Does there exist a set $F \subseteq E$ with $|F| \le k$ such that 
    the graph $G - F$ can be partitioned into at most $d$ clusters, 
    each having diameter at most $s$?
  \end{minipage}%
}
\vspace{3mm}

\begin{corollary}\label{cor:pw+d}
{\sc $s$-Club $d$-Cluster Edge Deletion} is W[1]-hard when parameterized by the 
combined parameter $\pw(G) + d$.
\end{corollary}

\begin{proof}
We modify the reduction used in the proof of Theorem~\ref{hardness-result} by 
a small change in the construction of the gadgets corresponding to the vertices 
of the independent set $I$ in the \textsc{Minimum Maximum Outdegree} instance 
$(G,w,r)$.

\smallskip
\noindent
\emph{Modified construction.}
Recall from Figure~6(b) that in the original reduction each vertex
$b_s\in I$ had its own gadget $H_{b_s}$ together with a private auxiliary
path between $x_0^{b_s}$ and $x_0^{*b_s}$.
In the modified construction, we replace all these auxiliary paths by a
single shared auxiliary path.

More precisely, we identify the vertices
$\{x_0^{b_1},x_0^{b_2},\ldots,x_0^{b_{n'}}\}$ into one vertex $x_0^I$,
and identify
$\{x_0^{*b_1},x_0^{*b_2},\ldots,x_0^{*b_{n'}}\}$ into one vertex
$x_0^{*I}$.
We then connect $x_0^{*I}$ and $x_0^I$ by a single auxiliary path of
length $s-(\alpha r+2n+2)$.
All remaining parts of the gadgets $H_{b_s}$ remain unchanged; in
particular, each original spine now branches from the common vertex
$x_0^I$.
We refer to this gadget as $H_{b}$.
See Figure~7 for an illustration.

Observe that identifying the vertices into the common vertex $x_0^I$
does not increase the diameter beyond the bound enforced in the original
construction.
Indeed, every branch corresponding to a gadget $H_{b_s}$ still has
distance at most $\alpha r+2n$ from $x_0^I$, while the common auxiliary
path between $x_0^{*I}$ and $x_0^I$ has length
$s-(\alpha r+2n+2)$.
Thus every vertex of the merged gadget remains within bounded distance
from the central vertex $x_0^I$, and the same diameter argument as in
Theorem~\ref{hardness-result} continues to apply.

\begin{figure}[ht]
\centering
\begin{tikzpicture}[scale=0.6,
  x=1cm,y=1cm,
  v/.style={circle,draw,fill=white,inner sep=0pt,minimum size=3.2pt},
  ed/.style={line width=.8pt},
  reded/.style={very thick,red},
  lab/.style={font=\scriptsize}
]
\node[v] (xb10) at (0,-6) {};
\node[v] (xb11) at (4,-6) {};
\node[v] (xb12) at (8,-6) {};
\node[lab,above left=2pt] at (xb10) {$x_{0}^{I}$};
\node[lab,below=2pt] at (xb11) {$x_{1}^{b_{1}}$};
\node[lab,below=2pt] at (xb12) {$x_{2}^{b_{1}}$};

\node[v] (y0b1)  at (0,0) {};
\node[v] (yb1a1) at (2,-5) {};
\node[v] (yb1a2) at (6,-5) {};
\node[lab,left=3pt] at (y0b1)  {$x_{0}^{*I}$};
\node[lab,above=3pt] at (yb1a1) {$y_{a_1}^{b_{1}}$};
\node[lab,above=3pt] at (yb1a2) {$y_{a_2}^{b_{1}}$};

\draw[reded] (y0b1)--(xb10);
\draw[reded] (xb11)--(xb10);
\draw[reded] (xb12)--(xb11);
\draw[ed] (xb10)--(yb1a1);
\draw[ed] (xb11)--(yb1a1);
\draw[ed] (xb11)--(yb1a2);
\draw[ed] (xb12)--(yb1a2);

\node[v] (xb41) at (-4,-6) {};
\node[v] (xb42) at (-8,-6) {};
\node[lab,below=2pt] at (xb41) {$x_{1}^{b_{4}}$};
\node[lab,below=2pt] at (xb42) {$x_{2}^{b_{4}}$};

\node[v] (yb4a1) at (-2,-5) {};
\node[v] (yb4a2) at (-6,-5) {};
\node[lab,above=3pt] at (yb4a1) {$y_{a_1}^{b_{4}}$};
\node[lab,above=3pt] at (yb4a2) {$y_{a_2}^{b_{4}}$};

\draw[reded] (xb10)--(xb41);
\draw[reded] (xb42)--(xb41);
\draw[ed] (xb10)--(yb4a1);
\draw[ed] (xb41)--(yb4a1);
\draw[ed] (xb41)--(yb4a2);
\draw[ed] (xb42)--(yb4a2);

\node[v] (xb21) at (2.82,-7.82) {};
\node[v] (xb22) at (5.64,-10.64) {};
\node[lab,right=2pt] at (xb21) {$x_{1}^{b_{2}}$};
\node[lab,right=2pt] at (xb22) {$x_{2}^{b_{2}}$};

\node[v] (yb2a1) at (0.5,-7.82) {};
\node[v] (yb2a2) at (2.82,-10) {};
\node[lab,below=3pt] at (yb2a1) {$y_{a_1}^{b_{2}}$};
\node[lab,below=3pt] at (yb2a2) {$y_{a_2}^{b_{2}}$};

\draw[reded] (xb10)--(xb21);
\draw[reded] (xb22)--(xb21);
\draw[ed] (xb10)--(yb2a1);
\draw[ed] (xb21)--(yb2a1);
\draw[ed] (xb21)--(yb2a2);
\draw[ed] (xb22)--(yb2a2);

\node[v] (xb31) at (-3,-7.82) {};
\node[v] (xb32) at (-6,-10.64) {};
\node[lab,left=2pt] at (xb31) {$x_{1}^{b_{3}}$};
\node[lab,left=2pt] at (xb32) {$x_{2}^{b_{3}}$};

\node[v] (yb3a1) at (-0.5,-7.82) {};
\node[v] (yb3a2) at (-2.82,-10) {};
\node[lab,below=3pt] at (yb3a1) {$y_{a_1}^{b_{3}}$};
\node[lab,below=3pt] at (yb3a2) {$y_{a_2}^{b_{3}}$};

\draw[reded] (xb10)--(xb31);
\draw[reded] (xb32)--(xb31);
\draw[ed] (xb10)--(yb3a1);
\draw[ed] (xb31)--(yb3a1);
\draw[ed] (xb31)--(yb3a2);
\draw[ed] (xb32)--(yb3a2);

\begin{scope}[on background layer]
  \draw[
    rounded corners=3pt,
    draw=blue!60,
    fill=blue!20,
    fill opacity=0.18,
    line width=0.6pt
  ] (-9,-12) rectangle (9,1);
  \node[anchor=south west,font=\scriptsize,blue!60] at (9.5,-4.5) {$H_{b}$};
\end{scope}

\end{tikzpicture}
\caption{Modification of the gadgets corresponding to the independent-set vertices from Figure~6. 
Instead of having a separate auxiliary path for each gadget $H_{b_s}$, 
all vertices $\{x_0^{b_1},\ldots,x_0^{b_{n'}}\}$ are identified into a single vertex $x_0^I$, 
and all vertices $\{x_0^{*b_1},\ldots,x_0^{*b_{n'}}\}$ are identified into a single vertex $x_0^{*I}$. 
The resulting merged gadget is denoted by $H_b$. 
The common auxiliary path between $x_0^{*I}$ and $x_0^I$ is shown in red. 
Each branch incident with $x_0^I$ corresponds to the remaining spine structure of one original gadget $H_{b_s}$. 
For readability, every reinforced path is drawn as a single red edge.}
\label{mainfig}
\end{figure}

\smallskip
\noindent
\emph{Bounding the number of $2$-clubs.}  
Since there are no edges between distinct gadgets $H_{b_s}$ and $H_{b_{s'}}$ 
in the original construction, identifying their auxiliary endpoints does not 
introduce new edges between them.  
Consequently, in every feasible solution $G'-F$, all vertices of the sets 
$\{X^{b_s}\mid b_s \in I\}$ now belong to the same connected component, say $C_I$.
Each gadget $H_{a_t}$ corresponding to $a_t \in C$ continues to form its 
own component $C_{a_t}$ exactly as before.  
Hence the number of connected components (or $2$-clubs) in $G'-F$ satisfies
$d = \vc(G) + 1$,
which is bounded by a function of the vertex cover size of the source instance.

\smallskip
\noindent
\emph{Pathwidth bound.}  
The pathwidth of the constructed graph $G'$ remains bounded by the same function 
as before.  
In the original construction, each gadget could be placed in the path decomposition 
within a bounded number of consecutive bags.  
The identification of the endpoints 
$x_0^{b_s}$ and $x_0^{*b_s}$ into two common vertices $x_0^I$ and $x_0^{*I}$ 
does not increase the width: removing $x_0^I$ again separates all gadgets 
$\{H_{b_s}\}$ exactly as in the original argument.  
Thus, $\pw(G')$ remains bounded by a function of $\pw(G)$.

\smallskip
\noindent
\emph{Preservation of correctness.}  
The correctness arguments of Theorem~\ref{hardness-result} continue to hold 
verbatim.  
In particular:
\begin{enumerate}
    \item All vertices of each $X^{u}$ lie in a single component $C_u$.
    \item Distinct spines corresponding to vertices in the vertex cover $C$
    induce distinct components, and all vertices belonging to the independent-set
    gadgets jointly form one additional component $C_I$.
    \item For each edge $\{u,v\} \in E(G)$, the set $F$ must contain 
    either the two unique edges incident with $y^{u}_{v}$ whose other 
    endpoints lie in $X^{u}$, or the two unique edges incident with 
    $y^{v}_{u}$ whose other endpoints lie in $X^{v}$.  
    If neither pair were deleted, the reinforced path between $y^{u}_{v}$ 
    and $y^{v}_{u}$ would remain, merging $C_u$ and $C_v$ into a component 
    of diameter greater than $s$, contradicting feasibility.
\end{enumerate}
Since $|F| \le 2|E(G)|$, exactly one such pair is deleted for each 
$\{u,v\} \in E(G)$, defining a unique orientation of $G$ consistent with 
the solution to the source \textsc{MMO} instance.  
The modified identification does not affect this argument, because the 
gadgets $H_{b_s}$ never connect to each other directly, and the reinforced 
connections remain confined to pairs $\{H_u,H_v\}$ for $\{u,v\}\in E(G)$.

\smallskip
\noindent
\emph{Conclusion.}  
The reduction remains parameter-preserving and computable in polynomial time, 
and the correctness of the orientation–deletion correspondence holds unchanged.  
Therefore, {\sc $s$-Club Cluster Edge Deletion} is W[1]-hard when parameterized 
by $\pw + d$.
\end{proof}

\section{FPT under Treedepth, Cluster Vertex Deletion Number, and Neighbourhood Diversity}

In this section, we establish fixed-parameter tractability of the
{\sc $s$-Club Cluster Edge Deletion} problem under three structural
parameters of the input graph: treedepth, cluster-vertex-deletion
number, and neighbourhood diversity. Recall that treedepth measures
the hierarchical depth of a graph, cluster-vertex-deletion number is
the minimum number of vertices whose removal leaves a disjoint union
of cliques, and neighbourhood diversity measures the number of distinct
vertex neighbourhood types in the graph.

\subsection{FPT under Treedepth}

\begin{theorem}
{\sc $s$-Club Cluster Edge Deletion} can be solved in time $\mathcal{O}\!\left(2^{\,2^{\,\mathcal{O}(k^3)}} \cdot n^2\right)$,
and is therefore fixed-parameter tractable when parameterized by the treedepth~$k$ of the input graph.
\end{theorem}

\begin{proof}
It was shown in~\cite{MONTECCHIANI2023114041} that 
{\sc $s$-Club Cluster Edge Deletion} admits a dynamic programming algorithm 
running in time 
\(\mathcal{O}\!\left(2^{2^{\mathcal{O}(\operatorname{tw}(G)^2 \log s)}} \cdot n^2\right)\), 
parameterized by $s+\operatorname{tw}(G)$, where $\operatorname{tw}(G)$ 
is the treewidth of $G$.

Let $k = \td(G)$ denote the treedepth of $G$.  
By a result of Nešetřil and Ossona de Mendez~\cite[Chapter~6]{10.5555/2230458}, 
every connected graph of treedepth at most $k$ has diameter at most $2^k - 1$.  
Moreover, $\operatorname{td}(G) \le \operatorname{tw}(G) + 1$ holds for all graphs.
We analyze two cases:

\smallskip
\emph{Case 1: $s \ge 2^k$.}  
Then every connected component of $G$ already has diameter at most $s$, 
so $G$ itself is an $s$-club cluster, and the empty set is an optimal solution.  
Hence the instance can be solved trivially in polynomial time.

\smallskip
\emph{Case 2: $s < 2^k$.}  
We invoke the above dynamic programming algorithm with parameter 
$s + \operatorname{tw}(G) \le 2^k + (k-1)$.  
Substituting these bounds into the running time expression yields:
\[
\mathcal{O}\!\left(
  2^{\,2^{\,\mathcal{O}\!\bigl(k^2 \log (2^k-1)\bigr)}} \cdot n^2
\right)
 = \mathcal{O}\!\left(
  2^{\,2^{\,\mathcal{O}(k^3)}} \cdot n^2
\right).
\]
This is an FPT running time depending only on $k$.
Hence, {\sc $s$-Club Cluster Edge Deletion} is fixed-parameter tractable 
when parameterized by the treedepth of the input graph.
\end{proof}

\subsection{FPT under Cluster Vertex Deletion}

Having established tractability under treedepth, we now turn our attention
to the parameter cluster vertex deletion number. Recall that a cluster
vertex deletion set is a set of vertices whose removal turns the graph into
a cluster graph (a disjoint union of cliques). We also recall that two
vertices are called true twins if they have identical closed neighborhoods,
and false twins if they have identical open neighborhoods.

Before presenting our FPT result for {\sc $s$-Club Cluster Edge Deletion}
parameterized by cluster vertex deletion number, we first establish a useful
structural lemma concerning the distribution of twin classes across the
connected components of a solution. Informally, the lemma shows that twins
can be consolidated into essentially one component, which will later allow
us to bound the interaction between solution components and twin classes in
the kernelization arguments.

\begin{lemma}[Twin consolidation]\label{lem:twin-consolidation}
Let $s\ge 2$ and let $F$ be an optimal $s$-club cluster edge-deletion set of a
graph $G$. Let $T$ be a twin class in $G$.  
Then there exists another optimal solution $F'$ such that
in $G-F'$ all but at most one connected component contain at most one vertex of $T$.
\end{lemma}

\begin{proof}
Let $F$ be any optimal solution. 
Let the components of $G-F$ be $C_1,\dots,C_\ell$.  
Fix an arbitrary vertex $t\in T$. Set $\gamma_i := |N_G(t)\cap V(C_i)|$ for each $i$.
Choose $i^\star$ maximizing $\gamma_i$.

\smallskip
\proofpara{Removing redundant twins}  
Suppose $C$ is a component containing two twins $t,t'\in T$.  
Since $N(t)=N(t')$ (for false twins) or $N[t]=N[t']$ (for true twins),
any shortest path in $C$ that uses $t'$ can be rerouted through $t$
without increasing its length.  
Hence deleting $t'$ does not increase distances, and
$\diam(C-\{t'\})\le \diam(C)\le s$.  
Iterating, we keep only one vertex of $T$ in each component.

\smallskip
\proofpara{Adding twins elsewhere}  
Re-inserting deleted twins into a component never increases its diameter.
Let $C$ be a component with a surviving twin $t\in T$, and insert a new
copy $t'$ with $N(t')=N(t)$ (or $N[t']=N[t]$ in the true-twin case).  
For every $x\in V(C)$ we have 
$\dist_{C\cup\{t'\}}(t',x)=\dist_{C}(t,x)$,
and for true twins also $\dist(t,t')=1$.  
For false twins, if $N_C(t)\neq\emptyset$, then $t$ and $t'$ share a common
neighbor in $C$, giving $\dist(t,t')=2$; if $N_C(t)=\emptyset$, the new vertex
$t'$ is isolated and forms its own singleton component.  
Hence, in all cases (for $s\ge2$ when $T$ is a false-twin class),
$\diam(C\cup\{t'\})=\diam(C)\le s$.

\smallskip
\proofpara{Choice of target component} 
Recall that, for a fixed representative $t\in T$, we defined
$\gamma_i := |N_G(t)\cap V(C_i)|$
for every component $C_i$ of $G-F$, and chose $i^\star$ maximizing $\gamma_i$.
By Step (1) we may collect all but one twin from each component into a set
$R\subseteq T$. By Step (2) we may reinsert $R$ into any component already
containing a twin without increasing its diameter.  
If a vertex $t'\in R$ is placed in $C_j$, then $F$ must delete at least
$\deg_G(t)-\gamma_j$ edges.  
Placing all of $R$ into $C_{i^\star}$ instead costs at most
$\deg_G(t)-\gamma_{i^\star}$ per vertex.  
Since $\gamma_{i^\star}\ge \gamma_j$ for all $j$, this does not increase the
number of deletions. Hence $|F'|\le |F|$.
Therefore $F'$ is also optimal, and all but one connected component contain at most one
vertex of~$T$.

\smallskip
\noindent\emph{Remark.}
The assumption $s\ge 2$ is essential for false twins.  
If $s=1$, adding a false twin $t'$ to a component already containing its twin $t$
creates a pair at distance~2 (via a common neighbor), thus violating the
1-club property.  
For true twins, however, the distance between any two members of $T$ is~1,
so the lemma also holds for $s\ge 1$
\end{proof}

Now, we show that {\sc $s$-Club Cluster Edge Deletion} is FPT when parameterized by the cluster vertex deletion number of the input graph.

\begin{theorem}\label{CVD}
{\sc $s$-Club Cluster Edge Deletion} is FPT when
parameterized by the cluster vertex deletion number of the input graph.
\end{theorem}

\begin{proof}
Let \(X \subseteq V(G)\) be a cluster vertex deletion set of size \(k\).

\proofpara{Bounding the Diameter}
It is known that if \(\operatorname{cvd}(G) \le k\), then every connected
component of \(G\) has diameter at most \(3k+1\).
Hence, if \(s > 3k+1\), the instance is trivially a yes-instance,
since no edge deletions are required, and we may assume \(s \le 3k+1\).

\proofpara{Graph Transformation via Twin Consolidation}
We now describe how to transform the input instance \((G, s, k)\)
into an equivalent \emph{weighted} instance \((G', s, k)\) on which the
dynamic programming algorithm of~\cite{MONTECCHIANI2023114041} can be
applied.
For every clique \(C\) in \(G - X\),
partition the vertices of \(C\) into at most \(2^{k}\) twin classes
according to their neighborhoods in \(X\).
Since the vertices within each class are true twins, deleting any subset
of them affects only the multiplicity of edges to vertices outside the
class.

\smallskip
\noindent
By Lemma~\ref{lem:twin-consolidation}, there exists an optimal solution $F$ such that all but at most one connected component of $G-F$ contain at most one vertex from each true twin class $T$. Hence, among vertices of $T$, at most one component may require multiple representatives of $T$, while every other component needs at most one representative. Since any solution deletes at most $k$ edges, keeping more than $k+1$ vertices of $T$ is unnecessary: one may keep up to $k$ explicit representatives together with one additional weighted representative encoding all remaining equivalent vertices.
Therefore, for each true twin class \(T\), the presence of more than \(k+1\)
vertices is redundant, as deleting or keeping additional equivalent vertices
does not change the existence or size of an optimal solution.
Hence, if \(|T| > k+1\), we remove all but \(k\) vertices of \(T\) and
replace the remaining \(|T| - k\) surplus vertices by a single
representative vertex.
Each edge incident to this representative vertex is assigned a weight equal
to the number of vertices it replaces, ensuring that the total deletion cost
correctly reflects the original multiplicity of identical edges.
This operation preserves all adjacencies and pairwise distances between
distinct twin classes and produces an equivalent instance for computing an
optimal \(s\)-club cluster edge-deletion set.

More formally, for each twin class \(T\subseteq C\), we apply the following reduction rule: 
\begin{itemize}
    \item If \(|T| \le k+1\), we keep all vertices of \(T\) unchanged.
    \item If \(|T| = k + \delta\) for some \(\delta > 0\),
    we remove the extra \(\delta\) vertices from \(T\)
    and replace them by a single representative vertex \(t'\).
    For every vertex \(v \in N_G(T)\),
    we introduce an edge \(\{t',v\}\) of \emph{weight} \(w(\{t',v\}) = \delta\),
    while all other edge weights are set to~1.
    The neighborhood of \(t'\) is exactly \(N_G(T)\).
\end{itemize}

Applying the above transformation independently to every clique \(C\) of \(G-X\),
we obtain a weighted graph \(G'\). Since each clique \(C\) is partitioned into at
most \(2^k\) twin classes, and each twin class contributes at most \(k+1\)
vertices after the reduction, every clique of \(G'-X\) contains at most
\(2^k(k+1)\) vertices. Moreover, all edge weights in \(G'\) are positive integers.
Finally, \(G'\) admits a path decomposition of width at most
\(k + (k+1)2^k\): for each clique of \(G'-X\), we introduce one bag containing
\(X\) together with all vertices of the corresponding reduced clique.

\proofpara{Equivalence of Instances}
We claim that \((G, s, k)\) is a yes-instance if and only if
\((G', s, k)\) is a yes-instance.

($\Rightarrow$) Suppose \(F \subseteq E(G)\) is an optimal edge deletion
set of size at most \(k\) such that \(G-F\) is an \(s\)-club cluster graph.
For each deleted edge \(\{u,v\}\) in \(G\), we delete the corresponding
edge in \(G'\).

Since edges incident with consolidated vertices in \(G'\)---that is,
weighted representative vertices replacing multiple equivalent vertices
from a twin class---represent multiple identical edges in \(G\),
the total deletion cost in \(G'\) equals the number of such edges in \(G\),
and thus \(\sum_{e\in F} w(e) \le k\).
Moreover, since consolidation preserves adjacency relations and
distances among distinct twin classes, the resulting \(G'-F'\) remains
an \(s\)-club cluster graph.

($\Leftarrow$) Conversely, let \(F'\) be an optimal solution in \(G'\)
of total cost at most \(k\).
For each weighted edge \(e=\{u,v\}\) with \(w(e)=\delta\) that is deleted,
we delete \(\delta\) parallel edges (i.e., edges incident to the
corresponding \(\delta\) twins) in \(G\).
By expanding all consolidated vertices, we recover an edge deletion set
\(F\subseteq E(G)\) of size \(|F| \le k\)
such that \(G-F\) is an \(s\)-club cluster graph.
Therefore, the two instances are equivalent.

\proofpara{Applying the Weighted Dynamic Program.}
We now argue that the dynamic programming algorithm of~\cite{MONTECCHIANI2023114041}
can be applied directly to \(G'\).
Recall that, for every bag \(X_i\) of the nice tree decomposition, the algorithm stores a set of \emph{solutions}, each represented by a tuple
\[
S^l_i = \langle \partial \mathcal{C}^l_i, D^l_i, H^l_i, Q^l_i, \mu^l_i\rangle.
\]
Here:
\begin{itemize}
    \item \(\partial \mathcal{C}^l_i\) is the family of boundaries of the potential \(s\)-clubs intersecting \(X_i\);
    \item \(D^l_i\) is the set of distance tables \(D(\partial C)\) that record, for each pair \(a,b\in \partial C\), the truncated distance \(d_{G[C]}(a,b)\) capped at \(s\);
    \item \(H^l_i\) stores, for each equivalence class of vertices in \(\mathrm{int}(C)\), their distances to vertices in \(\partial C\);
    \item \(Q^l_i\) is the set of pending \emph{requests} \(R_{wz}\) over \(\partial C\), each encoding potential paths of length at most \(s-2\) that could make a pair of interior vertices \(w,z\) satisfy the distance constraint;
    \item finally, \(\mu^l_i\)is the \emph{edge-counter} representing the number of edges deleted in the partial solution corresponding to the subgraph \(G_i\).
\end{itemize}

In the weighted setting, these components remain unchanged,
except that \(\mu^l_i\) is reinterpreted as the \emph{total edge weight}
of deleted edges, namely,
\[
\mu^l_i = \sum_{e\in E_i \setminus P^l_i(E_i)} w(e).
\]
Whenever a deleted edge \(e\) appears in a transition,
the algorithm updates
\(\mu^l_i \leftarrow \mu^l_i + w(e)\)
instead of incrementing by one.
Since the recurrence relations and feasibility conditions depend only on
the structural components
\((\partial \mathcal{C}^l_i, D^l_i, H^l_i, Q^l_i)\),
the correctness proof of Lemma~2 in~\cite{MONTECCHIANI2023114041}
remains valid verbatim.

Therefore, the dynamic program computes an optimal weighted solution on \(G'\),
which by the above equivalence corresponds to an optimal unweighted solution
on \(G\).

\smallskip
\noindent
We now analyze the running time more precisely.
The dynamic programming algorithm of~\cite{MONTECCHIANI2023114041} runs in
\(\mathcal{O}\!\bigl(2^{\,2^{\,\mathcal{O}(\mathrm{tw}(G')^{2}\log s)}}\cdot n^{2}\bigr)\)
time, where \(\mathrm{tw}(G')\) is the treewidth of the input graph
and \(s\) is the cluster diameter bound.
For our transformed graph \(G'\), we have
$\mathrm{tw}(G') \le k + (k+1)2^{k}
\quad\text{and}\quad
s \le 3k+1$.
Since \(k+(k+1)2^{k} \le (k+2)2^{k}\),
we can further simplify the exponent to obtain
$(k+(k+1)2^{k})^{2}\log(3k+1)
=\mathcal{O}\!\bigl(4^{k}\,k^{2}\log k\bigr)$,
and hence the total running time is
$T(k,n)
=\mathcal{O}\!\left(
2^{\,2^{\,\mathcal{O}\!\left(4^{k}\,k^{2}\log k\right)}}\cdot n^{2}
\right)$.

\noindent
Thus, {\sc $s$-Club Cluster Edge Deletion} is fixed-parameter tractable when
parameterized by the cluster vertex deletion number \(k\),
with the above double-exponential dependence in \(k\) and polynomial dependence in \(n\).
\end{proof}

\subsection{Algorithm under Neighbourhood Diversity Parameterization}

In this section, we show that {\sc $s$-Club Cluster Edge Deletion} is fixed-parameter tractable when parameterized by the neighbourhood diversity of the input graph. 
For the special case $s=1$, corresponding to {\sc Cluster Edge Deletion}, the problem is known to admit an algorithm running in $2^{2^{\mathcal{O}(\nd)}} \cdot n^{\mathcal{O}(1)}$ time~\cite{10.1007/978-3-031-27051-2_31}. 
We therefore focus on the case $s \ge 2$. 
Our approach is fundamentally different: rather than using the linear-integer programming formulation employed for the $s=1$ case, we rely on the structural characterization provided by the twin-consolidation lemma to bound the number of relevant component partitions. 
Nevertheless, the resulting algorithm achieves the same asymptotic dependence on the parameter as in the $s=1$ case, up to constant factors.

\begin{theorem}\label{thm:nd-fpt}
Let $G$ be a graph with neighborhood diversity $\nd(G)=k$ and let $s\ge 2$.
Then {\sc $s$-Club Cluster Edge Deletion} can be solved in
$2^{\,k(2^k+1)}\cdot \mathcal{O}(nm+n^2)$
time. In particular, the problem is fixed-parameter tractable when
parameterized by $\nd(G)$.
\end{theorem}

\begin{proof}
We exploit the structure induced by the neighborhood-diversity partition of \(G\).
At a high level, we classify connected components of a solution according to
which twin classes they intersect. We then show that only a bounded number of
such component types need to be considered, which allows us to enumerate all
feasible clusterings in FPT time.

\proofpara{Neighborhood-diversity structure}
Let $\mathcal{C}=\{T_1,\ldots,T_k\}$ be a neighborhood-diversity partition of \(V(G)\):
for each \(i\neq j\), the bipartite graph between \(T_i\) and \(T_j\)
is either complete or empty.
Each $T_i$ is therefore either a clique (true-twin class) or an independent set
(false-twin class).
Let the \emph{type graph} $H$ be the graph on vertex set $[k]=\{1,\dots,k\}$
where $\{i,j\}\in E(H)$ if and only if $G$ is complete between $T_i$ and $T_j$.

\smallskip
We now reason about how the structure of $H$ constrains the components that may
appear in an optimal clustering after edge deletions.

\proofpara{Footprints and equivalence of components}
Given an edge-deletion set $F\subseteq E(G)$ producing the clustering $G-F$,
each connected component $C$ of $G-F$ determines its \emph{footprint}
$\mathrm{foot}(C):=\{\,i\in[k]\;|\;T_i\cap V(C)\neq\emptyset\,\}$.
Two components $C_1$ and $C_2$ are called \emph{equivalent} if
$\mathrm{foot}(C_1)=\mathrm{foot}(C_2)$.

\smallskip
We next show that whenever two components share the same footprint, it is safe
to merge them without violating the $s$-club property.

\proofpara{Merging equivalent components}
Assume $C_1,C_2$ are equivalent and that for every
$i\in\mathrm{foot}(C_1)$ either $T_i$ is a clique or
$\mathrm{foot}(C_1)$ contains a neighbor of $i$ in $H$.
Let
$F' := F \setminus E_G(C_1,C_2)$,
that is, we restore all edges between $V(C_1)$ and $V(C_2)$ while
keeping all other deletions unchanged.
Let $C$ denote the connected component of $G-F'$ containing
$V(C_1)\cup V(C_2)$.

We claim that $\diam(C)\le s$ provided $\diam(C_1)\le s$ and $\diam(C_2)\le s$.
Let $u\in T_{i_u}\cap V(C)$ and $v\in T_{i_v}\cap V(C)$, and let
$S:=\mathrm{foot}(C_1)=\mathrm{foot}(C_2)$.
Consider in $H[S]$ a shortest $i_u$–$i_v$ path
$i_u=\sigma_0,\sigma_1,\ldots,\sigma_\ell=i_v$
(with $\ell=0$ if $i_u=i_v$).
Since $G$ is complete between $T_{\sigma_t}$ and $T_{\sigma_{t+1}}$ for every $t$,
any choice of vertices $x_t\in T_{\sigma_t}$ yields a valid path
$x_0x_1\cdots x_\ell$ in $G$.
Both $C_1$ and $C_2$ contain at least one vertex from each $T_{\sigma_t}$,
so such a path lies entirely within $V(C)$.
Hence $\dist_C(u,v)\le \ell = \dist_{H[S]}(i_u,i_v)$.
Finally, if $i_u=i_v$ and $T_{i_u}$ is a false-twin class, the assumption on
$S$ ensures that $i_u$ has a neighbor in $H[S]$, yielding
$\dist_C(u,v)\le 2$.
In all cases, $\dist_C(u,v)\le \diam(H[S])\le s$, and thus merging preserves
the $s$-club property.

\smallskip
Having established that equivalent components can be merged safely,
we now use this to limit the number of distinct components that may appear.

\proofpara{Bounding the number of components}
The merging argument implies that in an optimal solution we may assume there
is at most one component per footprint $S\subseteq [k]$,
except possibly for \emph{degenerate} footprints $S=\{i\}$
where $T_i$ is a false-twin class with no neighbor in $H$.
Such components are isolated singletons, trivially forming $s$-clubs for $s\ge 2$.
Consequently, the number of distinct footprints realized by components of
$G-F$ is at most $2^k$, and for each non-degenerate footprint there is at most
one corresponding component.

\smallskip
To further restrict the structure of these components, we next apply the
twin-consolidation lemma.

\proofpara{Twin consolidation within footprints}
By Lemma~\ref{lem:twin-consolidation}, we may further assume that in $G-F$
all but at most one component contain at most one vertex of each class $T_i$;
a single designated component (if any) may contain the remaining vertices of
$T_i$ without violating the $s$-club property.

\smallskip
With this structural restriction in place, we can now enumerate all admissible
ways of distributing the twin classes among the possible component footprints.

\proofpara{Enumerating admissible partitions}
Let $C\le 2^k$ denote the number of non-degenerate footprints to be realized.
For a fixed class $T_i$, we distribute its vertices among these $C$ components
as follows:
(i) choose the designated component (or none), and
(ii) for each remaining component, decide whether it receives a single
representative of $T_i$ or not.
Thus the number of admissible distributions per class is at most
$C\cdot 2^{C}$, and the total number of global distributions (over $k$ classes)
is bounded by
$\bigl(C\cdot 2^{C}\bigr)^{k}
\;\le\;
\bigl(2^{k}\cdot 2^{2^{k}}\bigr)^{k}
\;=\;
2^{\,k(2^{k}+1)}$.

\smallskip
We now explain how each enumerated partition corresponds to a concrete
edge-deletion solution.

\proofpara{From a partition to a solution and verification}
For each guessed distribution, we instantiate the components by choosing
arbitrary representatives in each $T_i$ (since all are twins),
delete all edges whose endpoints lie in different components,
and verify that every component has diameter at most $s$.
This verification uses BFS from each vertex within its component and takes
$\mathcal{O}(nm)$ time overall, plus $\mathcal{O}(n^2)$ bookkeeping for
component distances, for a total of $\mathcal{O}(nm+n^2)$ per trial.

For the decision version with deletion budget $\kappa$, we count crossing edges
and accept iff this number is at most $\kappa$.
For the optimization version, we retain the partition yielding the smallest
number of crossing edges among all valid configurations.

\smallskip
Finally, we analyze the runtime and summarize the correctness argument.

\proofpara{Running time and correctness}
The enumeration contributes a factor of $2^{\,k(2^{k}+k)}$, and each partition
is verified in $\mathcal{O}(nm+n^2)$ time.
Correctness follows from:
(i) the merging argument, ensuring at most one component per non-degenerate footprint;
(ii) the twin-consolidation lemma, bounding how many copies of a class appear per component;
and (iii) the fact that the minimum deletion set realizing a fixed partition is
precisely the set of crossing edges.
Therefore, the algorithm runs in
$2^{\,k(2^{k}+1)}\cdot \mathcal{O}(nm+n^2)$ time and establishes fixed-parameter
tractability for the parameter $\nd(G)=k$.
\end{proof}

\section{No Polynomial Kernelization}
In this section, we show that {\sc $s$-Club Cluster Edge Deletion} does not admit a polynomial kernel 
when parameterized by the vertex cover number of the input graph.
Our proof proceeds via a polynomial parameter transformation 
from the classical {\sc Red--Blue Dominating Set} problem, 
defined in Section~\ref{problems section}.
It is known that {\sc RBDS}, parameterized by $|B|$, 
does not admit a polynomial kernel unless 
$\mathrm{NP}\subseteq \mathrm{coNP}/\mathrm{poly}$~\cite{fomin_lokshtanov_saurabh_zehavi_2019}.  
We strengthen this result by showing that the same lower bound 
holds even when all red vertices have equal degree in $B$.

\begin{lemma}\label{lem:rbds-regular}
{\sc Red--Blue Dominating Set}, restricted to instances
in which all red vertices have the same number of neighbours in the blue set,
does not admit a polynomial kernel when parameterized by $|B|$,
unless $\mathrm{NP}\subseteq \mathrm{coNP}/\mathrm{poly}$.
\end{lemma}
\begin{proof}
We provide a polynomial-parameter transformation from the general version of
{\sc Red--Blue Dominating Set (RBDS)}
to a restricted version in which all red vertices have the same number of
blue neighbors.

\proofpara{Starting instance}
Let $I=(G=(R\cup B,E),k)$ be an instance of {\sc RBDS},
where $R$ and $B$ are the sets of red and blue vertices, respectively.
We construct an equivalent instance
$I'=(G'=(R'\cup B',E'),k')$
such that every red vertex of $G'$ has the same number of blue neighbors.

\proofpara{Construction of $G'$}
We build a bipartite graph $G'=(R'\cup B',E')$ as follows.

\begin{enumerate}
    \item Create two disjoint copies of $B$, denoted $B_1$ and $B_2$.
    For every $u\in B$, let $u_1\in B_1$ and $u_2\in B_2$ denote its copies.
    Add two new blue vertices $x_1$ and $x_2$, and define $B' := B_1 \cup B_2 \cup \{x_1,x_2\}$.
    \item Add two new red vertices $y_1,y_2$ and set
    $R' := R \cup \{y_1,y_2\}$.
    \item For each $v\in R$ and $u\in B$:
    \begin{itemize}
        \item if $\{v,u\}\in E(G)$, then add $\{v,u_1\}\in E'$;
        \item otherwise (if $\{v,u\}\notin E(G)$), add $\{v,u_2\}\in E'$.
    \end{itemize}
    Thus each $v\in R$ is adjacent to exactly one copy of every $u\in B$,
    implying $\deg_{G'}(v)=|B|$.

    \item Add the following edges for the special vertices:
    \begin{itemize}
        \item connect $y_1$ to $x_1$ and to any arbitrary $(|B|-1)$ vertices of $B_2$;
        \item connect $y_2$ to $x_2$ and to the remaining vertices of $B_2$
              so that $B_2\subseteq N_{G'}(y_1)\cup N_{G'}(y_2)$.
    \end{itemize}
    Hence $\deg_{G'}(y_1)=\deg_{G'}(y_2)=|B|$.
\end{enumerate}

Set the parameter $k' := k+2$.
This completes the construction of $G'$, which can clearly be done in polynomial time.
Every red vertex in $R'$ has exactly $|B|$ blue neighbors, establishing the uniform-degree property.

\proofpara{(\textrm{If}) direction}
Assume that $I=(G,k)$ is a yes-instance.
We construct a red set 
$D' := D \cup \{y_1,y_2\}$
in $G'$ and claim that $D'$ dominates $B'$.

\begin{itemize}
    \item For every $u_1\in B_1$, since $u$ has a neighbor $r\in D$ with $\{r,u\}\in E(G)$,
          we have $\{r,u_1\}\in E'$, so $u_1$ is dominated by $r\in D\subseteq D'$.
    \item For every $u_2\in B_2$, if $u_{2}$ is dominated by $y_{1}$ or $y_{2}$, since
          $B_2 \subseteq N_{G'}(y_1)\cup N_{G'}(y_2)$.
    \item The vertices $x_1$ and $x_2$ are dominated by $y_1$ and $y_2$, respectively.
\end{itemize}
Therefore, all vertices in $B'$ are dominated by $D'$,
and $|D'| = |D|+2 \le k+2 = k'$. 
Hence $I'$ is a yes-instance.

\proofpara{(\textrm{Only if}) direction}
Conversely, suppose that $I'=(G',k')$ is a yes-instance,
i.e., there exists $D'\subseteq R'$ with $|D'|\le k'$ that dominates all vertices of $B'$.
We show that $I=(G,k)$ is also a yes-instance.

Since $x_1$ and $x_2$ each have degree one,
their unique neighbors $y_1$ and $y_2$ must belong to $D'$.
Consequently, $|D'\setminus\{y_1,y_2\}|\le k'-2=k$.

Observe that neither $y_1$ nor $y_2$ has neighbors in $B_1$,
so $B_1$ must be dominated by $D'\setminus\{y_1,y_2\}$.
By construction, the induced subgraph $G'[R,B_1]$ is isomorphic to $G$.
Hence the set $D := D'\setminus\{y_1,y_2\}\subseteq R$ dominates all of $B$ in $G$
and satisfies $|D|\le k$.
Therefore, $I=(G,k)$ is a yes-instance.

\proofpara{Parameter preservation and conclusion}
The size of the blue set in $G'$ is $|B'| = 2|B| + 2 = \mathcal{O}(|B|)$,
and the transformation runs in polynomial time.
Thus, a polynomial kernel for {\sc RBDS} restricted to uniform-degree instances
would yield one for general {\sc RBDS} parameterized by $|B|$,
contradicting known lower bounds unless 
$\mathrm{NP}\subseteq \mathrm{coNP}/\mathrm{poly}$.
\end{proof}

Building on Lemma~\ref{lem:rbds-regular}, 
we establish our main lower bound. 

\paragraph{Overview of the reduction (Theorem~\ref{thm:no-poly-kernel-vc})}
We reduce from \textsc{Red--Blue Dominating Set} (RBDS).
Given an instance $G = (R \cup B, E)$ and an integer $k$, we construct
a graph $G'$ with vertex sets $R, B$ together with additional sets
$T, P, Q, S$ and vertices $w, y, z$ as in Figure~\ref{fig:clean-rbds-gadget}.

The set $B$ is made into a clique, and together with $P, Q$ and the
vertices $w, y, z$, it forms a part of the graph where all distances
can be kept at most $2$. In particular, $P$ connects $w$ and $y$,
while $Q$ connects $y$ and $z$, ensuring that these vertices must lie
in the same connected component in any solution of size at most $k'$.
Also, every vertex of $B$ is adjacent to $y$ and to all vertices of $P$.

The set $T$ together with $S$ forms another part of the graph.
All vertices in $R$ are adjacent to all vertices of $T$, and $T$ is
complete to $S$. This guarantees that once a vertex of $T$ remains
connected to $S$, the entire set $T \cup S$ lies in a single component
of diameter at most $2$.

Thus, any feasible deletion set $F$ must separate the vertices
$B \cup \{w,y,z\}$ from the set $T$. This forces a partition of the
graph into two components, say $C_1$ containing $B$ and $C_2$
containing $T$.

Now consider the vertices of $R$. Each vertex $r \in R$ has edges
to $B$, to $T$, and also an edge to $z$. If $r$ is placed in $C_1$
with $B$, then all edges from $r$ to $T$ must be deleted.
If $r$ is placed in $C_2$ with $T$, then all edges from $r$ to $B$
and the edge $\{r,z\}$ must be deleted. Since these costs are fixed,
the deletion budget $k'$ ensures that at most $k$ vertices from $R$
can be placed in $C_1$.

Let $X \subseteq R$ be the set of vertices that remain in the same
component as $B$. Then $|X| \leq k$. We show that the diameter
constraint forces $X$ to dominate $B$: for every $b \in B$, there must
exist a vertex $r \in X$ adjacent to $b$, otherwise $b$ would be at
distance greater than $2$ from $z$ in its component.

Conversely, given a dominating set $X \subseteq R$ of size at most $k$,
one can delete exactly the edges between the two parts so that the
graph splits into two components $C_1$ and $C_2$, each of diameter at
most $2$.

Finally, the construction ensures that the vertex cover of $G'$ is
contained in $B \cup T \cup \{w,y,z\}$ and hence has size $O(|B|)$.
This completes the reduction.

\begin{theorem}\label{thm:no-poly-kernel-vc}
{\sc $s$-Club Cluster Edge Deletion} does not admit a polynomial kernel 
when parameterized by the vertex cover number of the input graph,
even when $s=2$, unless $\mathrm{NP}\subseteq \mathrm{coNP}/\mathrm{poly}$.
\end{theorem}
\begin{proof}
We give a polynomial parameter transformation from {\sc RBDS}.  
By Lemma~\ref{lem:rbds-regular}, we may assume that every red vertex has the same number \(d\) of neighbors in \(B\).
\proofpara{Construction of \(G'\) from \(G\)}
Let \((G=(R\cup B,E),k)\) be an {\sc RBDS} instance.  
Starting from \(G\), create \(G'\) by adding the following new vertex sets:
\begin{itemize}
  \item a set \(T\) of size \(|T|:=|B|+2\);
  \item three special vertices \(w,y,z\);
  \item three sets \(P,Q,S\) with \(|P|=|Q|=|S|=k'+1\) (where \(k'\) will be set below).
\end{itemize}

The edge set is given by
\[
\begin{aligned}
E(G') :=\;& E(G)
   \;\cup\; K(\{w,y\}, P\cup Q)
   \;\cup\; K(y, B)
   \;\cup\; K(z, R\cup Q) \\[2pt]
 & \;\cup\; K(B, P)
   \;\cup\; K(R, T)
   \;\cup\; K(T, S)
   \;\cup\; K(B)
   \;\cup\; \{\{w,z\}\}.
\end{aligned}
\]
where \(K(A,B)\) denotes all edges between disjoint sets \(A\) and \(B\),
and \(K(A)\) denotes that \(A\) induces a clique. Please refer to Figure~\ref{fig:clean-rbds-gadget} to understand the construction.

\begin{figure}[t]
\centering
\begin{tikzpicture}[
  set/.style={ellipse,draw,minimum width=12mm,minimum height=5mm,inner sep=1pt},
  dot/.style={circle,draw,fill=black,inner sep=0pt,minimum size=3pt},
  lbl/.style={font=\small},
  vc/.style={draw=orange!70!black,fill=orange!25,fill opacity=0.6,very thick},
  thick,>=stealth,scale=0.8
]

\node[set,vc] (B) at (-2.6,0) {};     \node[lbl,left]  at (B.west) {$B$};
\node[set,vc] (T) at (7.2,1.4) {};       \node[lbl,above] at (T.north) {$T$};
\node[set] (P) at (-0.8,1.6) {};       \node[lbl,above] at (P.north) {$P$};
\node[set] (Q) at (1.8,1.6) {};        \node[lbl,above] at (Q.north) {$Q$};
\node[dot,vc] (w) at (0.5,2.2) {};     \node[lbl,above] at (w.north) {$w$};
\node[dot,vc] (y) at (-0.1,0.6) {};    \node[lbl,below right=-2pt and -1pt of y] {$y$};
\node[dot,vc] (z) at (3.2,1.4) {};     \node[lbl,above] at (z.north) {$z$};

\node[set] (R) at (5.2,1.4) {};       \node[lbl,above] at (R.north) {$R$};
\node[set] (S) at (9.2,1.4) {};       \node[lbl,above] at (S.north) {$S$};

\draw (B) to[bend left=10] (P);
\draw (B) to[bend right=15] (y);
\draw (y) to[bend left=12] (P);
\draw (y) to[bend right=15] (Q);
\draw (w) to[bend left=10] (Q);
\draw (w) to[bend right=10] (P);
\draw (Q) to[bend left=10] (z);
\draw (w) .. controls (2,3) .. (z);

\draw (R) -- (T);
\draw (T) -- (S);

\draw (z)-- (R);

\end{tikzpicture}
\caption{Schematic view of the reduction gadget.  
For clarity, edges between the sets $B$ and $R$ are omitted; moreover, $B$ induces a clique. The orange vertices form a vertex cover.}
\label{fig:clean-rbds-gadget}
\end{figure}
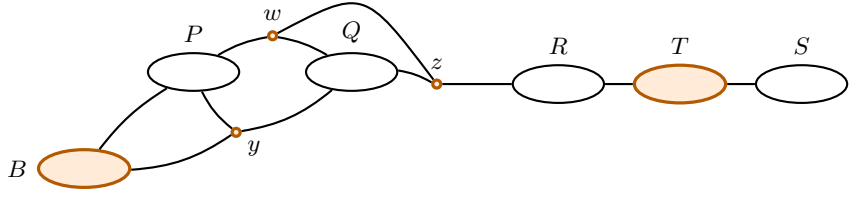

\proofpara{Parameter values}
We set
\[
s := 2
\qquad\text{and}\qquad
k' := k\cdot|T| + (n_R - k)\cdot(d+1)
= k(|B|+2) + (n_R - k)(d+1),
\]
where \(n_R := |R|\) and \(d := |N_B(r)|\) is the common number of blue neighbors for each \(r\in R\).

\proofpara{A small vertex cover of \(G'\)}
We claim that
$X' := B \cup T \cup \{w,y,z\}$
is a vertex cover of \(G'\). 
Indeed, by construction every edge of \(G'\) is incident to at least one vertex in \(X'\):
\begin{itemize}
  \item edges of $E(G)$, \(K(R,T)\), and \(K(T,S)\) have their endpoint in \(B\) or \(T\);
  \item edges of \(K(B,P)\) and the internal clique edges \(E(B)\) are incident with \(B\);
  \item edges of \(E(\{w,y\}, P\cup Q)\) are incident with \(w\) or \(y\);
  \item edges of $K(y,B)$ are incident with $\{y\}$;
  \item edges of \(K(z, R\cup Q)\) and the edge \(\{w,z\}\) are incident with \(z\).
\end{itemize}
Therefore, every edge of \(G'\) meets \(X'\), and thus \(X'\) is a vertex cover.
Its size satisfies
$\vc(G') \le |B| + |T| + 3 = 2|B| + 5$,
which is polynomial in \(|B|\).

\proofpara{Forward direction}
Assume \(X\subseteq R\) with \(|X|\le k\) dominates \(B\) in \(G\), i.e., every \(b\in B\) has a neighbor in \(X\).
Define a partition of \(V(G')\) into two vertex-disjoint sets:
\[
C_1 := B \cup P \cup Q \cup X \cup \{w,y,z\},
\qquad
C_2 := (R\setminus X) \cup T \cup S.
\]
Let \(F\) denote the set of all edges having one endpoint in \(C_1\) and the other in \(C_2\).

\proofpara{Bounding \(|F|\)}
The crossing edges are exactly:
\begin{itemize}
  \item for each \(x\in X\), all \(|T|\) edges \(\{x,t\}\) with \(t\in T\), contributing \(|X|\cdot|T|\);
  \item for each \(r\in R\setminus X\), the edge \(\{r,z\}\), contributing \(|R\setminus X|\);
  \item for each \(r\in R\setminus X\), its \(d\) edges \(\{r,b\}\in E\) to blue neighbors \(b\in B\subseteq C_1\), contributing \(|R\setminus X|\cdot d\).
\end{itemize}
No other edges cross the partition. Hence
\[
|F|
= |X|\cdot|T| + |R\setminus X|\cdot (d+1)
\le k\cdot|T| + (n_R-k)\cdot (d+1)
= k'.
\]

\proofpara{Diameter of \(C_2\)}
In \(G'[C_2]\), we have the complete bipartite graph \(K_{|T|,|S|}\) between \(T\) and \(S\), and each vertex in \(R\setminus X\) is adjacent to every vertex in \(T\). Thus:
\begin{itemize}
  \item any two vertices of \(R\setminus X\) have a path of length \(2\) via a common \(t\in T\);
  \item any two vertices of \(S\) have a path of length \(2\) via a common \(t\in T\);
  \item any \(r\in R\setminus X\) and \(s\in S\) have the path \(r\!-\!t\!-\!s\).
\end{itemize}
Hence \(\operatorname{diam}(G'[C_2])\le 2\).

\proofpara{Diameter of \(C_1\)}
Within \(C_1=B\cup P\cup Q\cup X\cup\{w,y,z\}\):
\begin{itemize}
  \item since \(B\) is a clique, all pairs of blue vertices are adjacent (distance 1);
  \item any two vertices of \(P\cup Q\cup\{w,y,z\}\) are within distance at most 2, 
        as \(w\) is adjacent to \(P\cup Q\) and \(z\) to \(Q\) and \(w\);
  \item for \(b\in B, p\in P\), the edge \(\{b,p\}\) gives distance 1;
  \item for \(b\in B, q\in Q\), \(b\!-\!y\!-\!q\) has length 2;
  \item for \(x,x'\in X\), \(x\!-\!z\!-\!x'\) has length 2;
  \item for \(x\in X, q\in Q\), \(x\!-\!z\!-\!q\) has length 2;
  \item for \(b\in B, x\in X\), the path \(b\!-\!x\) has length 1, if $b$ is not adjacent to $x$ then there exists another vertex $b'$ adjacent to $x$. In that case, take the path \(b\!-b'\!-\!x\) of length 2.
\end{itemize}
Therefore every pair of vertices in \(C_1\) is at distance at most 2, and
\(\operatorname{diam}(G'[C_1])\le 2\).

\smallskip
Hence we have constructed an edge-deletion set \(F\) with \(|F|\le k'\) such that each connected component of \(G'-F\) has diameter at most \(s=2\). This completes the forward direction.

\proofpara{Backward direction.}
Assume there exists an edge set \(F\subseteq E(G')\) with \(|F|\le k'\) such that
every connected component of \(G'-F\) has diameter at most \(s=2\).
We show that the original {\sc RBDS} instance \((G=(R\cup B,E),k)\) is a
YES-instance by constructing a red set \(X\subseteq R\) of size at most \(k\)
that dominates \(B\) in \(G\).

\proofpara{A robust ``anchor'' component containing \(B\cup\{w,y,z\}\)}
We claim that \(B\cup\{w,y,z\}\) lies in a single connected component of \(G'-F\).
Indeed:
\begin{itemize}
  \item For each fixed \(b\in B\), there are \(|P|=k'+1\) edge-disjoint
        \(b\)–\(y\) paths of the form \(b\!-\!p\!-\!y\) (one for each \(p\in P\)).
        Hence no set \(F\) with \(|F|\le k'\) can separate \(b\) from \(y\).
        Since \(B\) is a clique, all blue vertices remain mutually connected
        and each stays connected to \(y\).
  \item There are \(|Q|=k'+1\) edge-disjoint \(y\)–\(z\) paths of the form
        \(y\!-\!q\!-\!z\) (one for each \(q\in Q\)). Thus \(y\) and \(z\)
        cannot be separated by \(|F|\le k'\).
  \item Finally, \(w\) is adjacent to every \(p\in P\) and \(y\) to every
        \(p\in P\), so \(w\) and \(y\) are joined by \(|P|=k'+1\) edge-disjoint
        2-paths \(w\!-\!p\!-\!y\). Hence \(w\) remains in the same component
        as \(y\), and therefore as \(B\) and \(z\).
\end{itemize}
By transitivity, \(B\cup\{w,y,z\}\) is contained in a single component of \(G'-F\);
denote this component by \(C_1\).

\proofpara{A robust component containing \(T\) and at least one vertex of \(S\)}
Between any two distinct \(t,t'\in T\) there exist \(|S|=k'+1\) edge-disjoint
paths \(t\!-\!s\!-\!t'\) (one via each \(s\in S\)). Hence no \(|F|\le k'| \)
can separate vertices of \(T\) from each other, so \(T\) lies entirely in a
single component of \(G'-F\). Moreover, deleting all \(T\)–\(S\) edges would
require \(|T|\cdot|S|=(|B|+2)(k'+1)>k'\) deletions; thus at least one
vertex \(s^\star\in S\) remains adjacent to \(T\) in \(G'-F\). Denote by
\(C_2\) the component containing \(T\) (and some \(s^\star\)).

\proofpara{\(B\) and \(T\) must be in different components.}
Suppose for contradiction that some \(t\in T\) and some \(b\in B\) lie in the
same component of \(G'-F\). Any \(b\)–\(t\) walk must use an \(R\)-vertex
(because there is no edge from \(B\) or \(\{w,y,z,P,Q\}\) to \(T\)). Hence the
shortest \(b\)–\(t\) path in \(G'\) has length at least \(3\). This contradicts
the constraint \(\operatorname{diam}\le 2\). Therefore \(C_1\) and \(C_2\) are
distinct, and in particular \(B\subseteq V(C_1)\) while \(T\subseteq V(C_2)\).

\proofpara{Let \(X:=R\cap V(C_1)\); a lower bound on \(|F|\) in terms of \(|X|\)}
Because \(T\subseteq V(C_2)\) and every \(r\in R\) is adjacent to every \(t\in T\),
all \(|T|\) edges from each \(x\in X\) to \(T\) must be deleted to keep \(x\) out
of \(C_2\). Thus these contribute \(|X|\cdot|T|\) edges to \(F\).
Similarly, each \(r\in R\setminus X\subseteq V(C_2)\) has the edge \(\{r,z\}\)
(to \(z\in V(C_1)\)) and exactly \(d\) edges to its blue neighbors in \(B\subseteq V(C_1)\);
all \(d+1\) such edges must be deleted. Hence
\[
|F|\;\ge\; |X|\cdot|T| \;+\; |R\setminus X|\cdot(d+1)
 \;=\; k' \;+\; (|X|-k)\cdot\bigl(|T|-(d+1)\bigr),
\]
since \(k' = k|T| + (|R|-k)(d+1)\).

Because \(d=|N_B(r)|\le |B|\) for each \(r\in R\) and \(|T|=|B|+2\),
we have \(|T|-(d+1)=|B|-d+1\ge 1\). Therefore, if \(|X|=k+\ell\) with \(\ell\ge 1\),
then
\[
|F|\;\ge\; k' \;+\; \ell\cdot\bigl(|T|-(d+1)\bigr) \;\ge\; k'+\ell \;>\; k',
\]
contradicting \(|F|\le k'\). Hence necessarily \(|X|\le k\).

\proofpara{\(X\) dominates \(B\) (and \(|X|\le k\))}
Fix \(b\in B\). Since \(b,z\in V(C_1)\) and \(\operatorname{diam}(C_1)\le 2\),
we must have \(\operatorname{dist}_{C_1}(b,z)\le 2\). There is no edge
\(\{b,z\}\) in the construction, and there is no 2-path \(b\!-\!y\!-\!z\) (we did not
add \(\{y,z\}\)), nor a 2-path through \(P\) or \(Q\) (any such route uses
\(w\) or \(y\) and has length at least \(3\)). The only possible 2-path from
\(b\) to \(z\) is therefore \(b\!-\!x\!-\!z\) with \(x\in R\).
Because this path must lie inside \(C_1\), we conclude that \(x\in X\) and
\(\{b,x\}\in E\). Since \(b\) was arbitrary, every blue vertex has a neighbor
in \(X\). Combining with \(|X|\le k\), the set \(X\) is a red dominating set
of size at most \(k\) for \(B\) in \(G\).

\smallskip
We have shown that any solution \(F\) of size at most \(k'\) in \(G'\) yields
a red set \(X\subseteq R\) with \(|X|\le k\) that dominates \(B\) in \(G\).
This completes the backward direction.
\end{proof}

Theorem~\ref{thm:no-poly-kernel-vc} thus provides strong evidence 
that kernelization for {\sc $s$-Club Cluster Edge Deletion} is inherently difficult, 
even when the input graph is highly structured (small vertex cover) 
and the diameter bound $s$ is fixed.
We remark that this complements our positive FPT results: 
although the problem is fixed-parameter tractable when parameterized by the vertex cover number,
it does not admit a polynomial-size kernel unless a major collapse occurs in the polynomial hierarchy.

\begin{corollary}\label{cor:no-poly-kernel-vc-d2}
Even when restricted to \(s=2\) and \(d=2\), {\sc $s$-Club $d$-Cluster Edge Deletion}
does not admit a polynomial kernel when parameterized by the vertex cover number,
unless \(\mathrm{NP}\subseteq \mathrm{coNP}/\mathrm{poly}\).
\end{corollary}

\begin{proof}
This follows immediately from Theorem~\ref{thm:no-poly-kernel-vc}.
In the very same construction and proof of Theorem~\ref{thm:no-poly-kernel-vc},
every yes-instance produces exactly two diameter-\(\le 2\) components in \(G'-F\)
(the “anchor” component containing \(B\cup\{w,y,z\}\) and the “\(T\)-side” component),
so the bound \(d=2\) holds without any modification. The parameter \(\vc(G')\)
remains unchanged and polynomial in \(|B|\), hence the kernel lower bound
carries over verbatim.
\end{proof}

\section{Parameterized by $k$}

The current algorithmic landscape shows that {\sc $s$-Club Cluster Edge Deletion}
is fixed-parameter tractable when parameterized by the combined parameter $s+k$.
However, the complexity under the single parameter $k$ remains unresolved.

A major obstacle in designing an FPT approximation algorithm parameterized
only by $k$ while preserving the exact diameter bound is that
{\sc $s$-Club Cluster Edge Deletion} is not monotone under edge deletions.
Indeed, deleting additional edges may increase distances and destroy feasibility.
Hence the family of feasible solutions is not closed under inclusion,
which prevents the direct application of many known FPT-approximation techniques.

This motivates the study of relaxed formulations that allow a small
multiplicative violation of the diameter bound.
In particular, we ask whether one can obtain an FPT algorithm
parameterized by $k$ if a $(1+\epsilon)$-violation of the diameter is allowed.
We answer this affirmatively by presenting an FPT bicriteria approximation
scheme for graphs excluding long induced cycles.

The following lemma shows that in graphs with no long induced cycles,
deleting a small number of edges cannot increase distances too much,
provided the vertices remain connected.
Intuitively, every deleted edge on a shortest path can be bypassed
through a short alternative path, since otherwise the graph would contain
a long induced cycle.
This observation will be the key ingredient in the bicriteria approximation
algorithm of Theorem~\ref{thm:approx-club}.

\begin{lemma}\label{lem:dist-after-deletion}
Let $G$ be a graph in which every induced cycle has length at most $t-1$, where $t\ge 4$.
Let $u,v\in V(G)$ with $\dist_G(u,v)=s$, and let $F\subseteq E(G)$ be a set of at most $k$ edges such that $u$ and $v$ remain connected in $G-F$. 
Then
\[
\dist_{G-F}(u,v)\le s+k(t-3).
\]
\end{lemma}

\begin{proof}
Fix a shortest $u$--$v$ path $P$ in $G$ of length $s$.  
If $F$ does not intersect $E(P)$, then $P$ survives in $G-F$ and 
$\dist_{G-F}(u,v)=s$, proving the claim.  
Hence assume that some edges of $P$ are deleted.
Let $xy\in E(P)\cap F$.  
Since $u$ and $v$ remain connected in $G-F$, there exists a shortest $x$--$y$ path 
$Q$ in $G-F$ avoiding $xy$.  
The cycle $C=Q\cup\{xy\}$ is chordless, for otherwise a chord would yield a strictly shorter $x$--$y$ path.  
As $G$ has no induced cycle of length $\ge t$, we obtain $|C|\le t-1$.  
Hence $|Q|\le t-2$, so replacing $xy$ by $Q$ increases the path length by at most 
$(t-2)-1 = t-3$.
Applying this argument independently for each deleted edge on $P$, we can 
substitute each removed edge by an alternate path of length at most $t-2$.  
Since at most $k$ edges of $P$ are deleted, the total length of the resulting 
$u$--$v$ walk in $G-F$ is at most $s+k(t-3)$.  
Therefore $\dist_{G-F}(u,v)\le s+k(t-3)$, as claimed.
\end{proof}

\begin{theorem}\label{thm:approx-club}
Let \(t\ge 4\) be a fixed constant and let \(\epsilon>0\).
For an instance $I=(G,s,k)$ of {\sc $s$-Club Cluster Edge Deletion} on an $n$-vertex
graph $G$ with no induced cycle of length at least $t$, there exists an FPT
bicriteria approximation algorithm parameterized by $k$, running in time
\[
\max\!\left\{\left(\frac{k(t-3)}{\epsilon}+1\right)^{k}\! \cdot n^{\mathcal{O}(1)},\;
k^{\mathcal{O}(k^{3})}\!\cdot n^{\mathcal{O}(1)}\right\}.
\]
\end{theorem}

\begin{remark}
The algorithm has the following one-sided guarantee:
\begin{itemize}
    \item on any input, the algorithm either outputs an edge set
    $F$ with $|F|\le k$ such that every connected component of
    $G-F$ has diameter at most $(1+\epsilon)s$, or it outputs NO;

    \item whenever the algorithm outputs NO, the instance $I$
    is indeed a NO-instance for the exact problem.
\end{itemize}

In particular, on every YES-instance the algorithm returns a
$(k,(1+\epsilon)s)$-feasible solution; on NO-instances it may
either return NO or return such a bicriteria solution.
\end{remark}

\begin{proof}
We analyze two regimes depending on the value of $s$.

\smallskip
\emph{Case~1: $s \leq \tfrac{k(t-3)}{\epsilon}$.}  
In this case, we invoke the known FPT algorithm for {\sc $s$-Club Cluster Edge Deletion}
parameterized by $s+k$~\cite{MONTECCHIANI2023114041}.  
This algorithm runs in time $(s+1)^{k}\cdot n^{\mathcal{O}(1)}$ and solves the problem exactly.  
If $I$ is a YES-instance, it returns an optimal deletion set $F$ of size at most $k$.  
If no such set exists, it correctly outputs ``NO''.  
Hence, in this regime, the algorithm provides an exact decision and solution.

\smallskip
\emph{Case~2: $s > \tfrac{k(t-3)}{\epsilon}$.}  
Here, we construct a reduction to the {\sc Edge Multicut} problem.  
Given a graph \(G=(V,E)\), a set of terminal pairs 
\(\mathcal{T}=\{(s_1,t_1),\ldots,(s_r,t_r)\}\subseteq V\times V\),
and an integer \(k\), {\sc Edge Multicut} asks whether there exists 
\(F\subseteq E\) with \(|F|\le k\) such that each terminal pair 
\((s_i,t_i)\in\mathcal{T}\) lies in distinct connected components of \(G-F\).

From our instance \(I=(G,s,k)\), we define
\[
\mathcal{T} := \{\, (u,v)\in V(G)\times V(G)\;:\; \dist_G(u,v) > s\,\}.
\]
That is, every vertex pair originally at distance more than $s$ becomes a terminal pair.  
Observe that if $F$ is a feasible deletion set for {\sc $s$-Club Cluster Edge Deletion}, 
then $(u,v)$ must be separated in $G-F$ for every $(u,v)\in\mathcal{T}$, since edge deletions can only increase distances.  
Thus, every feasible solution for {\sc $s$-Club Cluster Edge Deletion} of size at most $k$ 
is also a feasible solution for {\sc Edge Multicut} on $(G,\mathcal{T},k)$.

We now apply the FPT algorithm for {\sc Edge Multicut}~\cite{doi:10.1137/140961808}, 
which runs in time $k^{\mathcal{O}(k^{3})}\cdot n^{\mathcal{O}(1)}$.  
If this multicut instance is infeasible, then so is the original instance (since 
any $s$-club clustering would induce a feasible multicut).  
Hence, in this case, the algorithm safely outputs ``NO.''  
If the multicut instance is feasible, we obtain a set $F$ with $|F|\le k$ 
that separates all pairs in $\mathcal{T}$.

\smallskip
\emph{Bounding the diameter.}  
By construction, $F$ separates all pairs that were at distance greater than $s$ in $G$.  
Therefore, for any $u,v$ that remain in the same component of $G-F$, we have
$\dist_G(u,v) \le s$.  
In graphs that exclude induced cycles of length at least $t$, 
Lemma~\ref{lem:dist-after-deletion} implies that deleting up to $k$ edges 
can increase the distance between any two vertices by at most $k(t-3)$.  
Hence, in every component of $G-F$, we have
$\dist_{G-F}(u,v) \le s + k(t-3)$.
Since $s > \tfrac{k(t-3)}{\epsilon}$, we obtain
\[\frac{s + k(t-3)}{s} \le 1 + \frac{k(t-3)}{s} \le 1 + \epsilon,\]
which shows that each component of $G-F$ has diameter at most $(1+\epsilon)\cdot s$.

\smallskip
\emph{Conclusion.}  
If the multicut solver declares infeasibility, the input instance $I$ is guaranteed to be a NO-instance for the exact {\sc $s$-Club Cluster Edge Deletion} problem.  
Otherwise, we obtain a deletion set $F$ with $|F|\le k$ such that every component of $G-F$ 
has diameter at most $(1+\epsilon)\cdot s$.  
Thus, the algorithm either outputs ``NO'' (soundly) or returns a bicriteria 
$(k,(1+\epsilon)s)$-feasible solution.  
The total running time is
$$\max\!\left\{
\Bigl(\tfrac{k(t-3)}{\epsilon}+1\Bigr)^{k}\!\cdot n^{\mathcal{O}(1)},\;
k^{\mathcal{O}(k^{3})}\!\cdot n^{\mathcal{O}(1)}
\right\}.$$
This completes the proof.
\end{proof}

\section{\texorpdfstring{$s$}{s}-Club Cluster Arc Deletion (Directed Variant)}
\label{sec:directed-variant}

In this section, we study the parameterized complexity of the directed
variant of the {\sc $s$-Club Cluster Edge Deletion} problem when parameterized
by the cutset size, that is, by the number $k$ of arcs allowed to be deleted.

We work with finite simple digraphs \(D=(V,A)\). For \(u,v\in V\), let
\(\dist_D(u,v)\) be the length of a shortest directed \(u\!\to\!v\) path
(\(+\infty\) if no such path exists). A (weakly) connected subgraph
\(H\) of \(D\) is \emph{\(s\)-reach-bounded} if for all ordered pairs \((u,v)\in V(H)\times V(H)\),
whenever \(v\) is reachable from \(u\) in \(H\) we have \(\dist_H(u,v)\le s\).
Equivalently, \(H\) has directed eccentricity at most \(s\) from every vertex
to all vertices it can reach.

\vspace{3mm}
\noindent\fbox{%
  \begin{minipage}{0.96\linewidth}
    \textsc{\(s\)-Club Cluster Arc Deletion}\\[2pt]
    \textbf{Input:} A digraph \(D=(V,A)\) and integers \(k,s\ge 1\).\\
    \textbf{Question:} Does there exist a set \(F\subseteq A\) with \(|F|\le k\) such that
    every \emph{weakly} connected component \(C\) of \(D-F\) is \(s\)-reach-bounded; i.e.,
    for all \(u,v\in V(C)\), if \(v\) is reachable from \(u\) in \(C\) then \(\dist_{C}(u,v)\le s\)?
  \end{minipage}
}
\vspace{3mm}

The above definition imposes a one-sided distance constraint: we only require that whenever a vertex \(v\) is reachable from a vertex \(u\) in a component \(C\), the directed distance \(\dist_C(u,v)\) is at most \(s\); mutual reachability is not necessary. A stricter formulation sometimes considered in the literature would require each component of \(D-F\) to be strongly connected with directed diameter at most \(s\). Our definition follows the “reachable implies bounded distance” interpretation, which generalizes naturally from the undirected case. Finally, when \(D\) is obtained by replacing every undirected edge \(\{u,v\}\) in a graph \(G\) by both arcs \((u,v)\) and \((v,u)\), this directed variant coincides with the undirected \textsc{\(s\)-Club Cluster Edge Deletion} problem.

We show that this problem is computationally intractable even under 
structural restrictions. In particular, we prove that
\textsc{$s$-Club Cluster Arc Deletion} is
\(\mathrm{W[1]}\)-hard when parameterized by \(k\),
even when the input digraph is acyclic. In particular, we prove the following theorem:

\begin{theorem}\label{thm:directed-w1hard}
\textsc{$s$-Club Cluster Arc Deletion} is \(\mathrm{W[1]}\)-hard when
parameterized by the cutset size \(k\),
even when restricted to directed acyclic graphs (DAGs).
\end{theorem}

\begin{proof}
We give a parameterized reduction from 
\textsc{Edge Multicut}, defined in Section~\ref{problems section}. 
This problem is known to be $\mathrm{W[1]}$-hard when parameterized by the size of the cutset, 
even on directed acyclic graphs~\cite{10.1007/978-3-642-31594-7_49}.
Let the input instance be $(G,\mathcal{T},k)$.

\proofpara{Topological ordering of terminals}
Let $T=\{t_{1},t_{2},\ldots,t_{p}\}$ denote the set of all distinct terminals appearing in $\mathcal{T}$.
Since $G$ is acyclic, we fix a topological ordering of $V(G)$ and assume, without loss of generality,
that the terminals are indexed according to this order; that is, for any $1\le i<j\le p$,
there is no directed path from $t_{j}$ to $t_{i}$ in $G$.
All arcs added in our construction will respect this ordering, preserving acyclicity.

\proofpara{Construction of $D'$}
We construct a digraph $D'=(V',A')$ and an integer $s$
such that $(G,\mathcal{T},k)$ is a yes-instance of
\textsc{Edge Multicut} if and only if $(D',k',s)$ is a yes-instance of
\textsc{$s$-Club Cluster Arc Deletion}.

For each terminal $t_i\in T$, we perform the following steps:

\begin{itemize}
    \item Introduce two new vertices
    $t_i^{\mathrm{in}}$ and $t_i^{\mathrm{out}}$.

\item Introduce two directed chains
$L_i=(t_i^{\mathrm{in}},\ell^i_1,\ell^i_2,\ldots,\ell^i_{\ell-1})$
and also the chain
$R_i=(r^i_1,r^i_2,\ldots,r^i_{\ell-1},t_i^{\mathrm{out}})$,
where all these vertices are fresh and pairwise distinct.

\item Add the directed chain
\[
t_i^{\mathrm{in}}
\to
\ell^i_1
\to
\ell^i_2
\to
\cdots
\to
\ell^i_{\ell-1}
\to
t_i.
\]

\item Add the directed chain
\[
t_i
\to
r^i_1
\to
r^i_2
\to
\cdots
\to
r^i_{\ell-1}
\to
t_i^{\mathrm{out}}.
\]

    \item Each arc $(u,v)$ appearing on these chains is replaced by
    $k+1$ internally vertex-disjoint directed paths of length two from
    $u$ to $v$.
    We refer to such a gadget as a \emph{reinforced step}.
    This ensures that deleting at most $k$ arcs cannot disconnect
    the endpoints of the step.

    \item For every ordered pair $(t_i,t_j)\notin\mathcal{T}$ with $i<j$,
    add all arcs from vertices of $L_i$ to vertices of $R_j$; that is,
    for every $x\in L_i$ and every $y\in R_j$, add the arc $(x,y)$.

    \item For each $(t_i,t_j)\in\mathcal{T}$, add no such arcs.

    \item Retain all original arcs of $G$.
\end{itemize}

This completes the construction of $D'$. Refer to fig~\ref{fig:directed-construction} for example of the construction in Theorem~\ref{thm:directed-w1hard}.

\begin{figure}[ht]
\centering
\begin{tikzpicture}[
    scale=0.72,
    v/.style={circle,fill=black,inner sep=1.2pt},
    arc/.style={->,>=stealth,thick},
    bluearc/.style={->,>=stealth,thick,blue!55},
    box/.style={draw,thick},
    midarrow/.style={
        postaction={decorate},
        decoration={
            markings,
            mark=at position 0.5 with {\arrow{stealth}}
        }
    },
    font=\small
]

\draw[box] (-8,-1) rectangle (8,1);
\node[left] at (-8,0) {$G$};

\foreach \x/\lab in {
-7/{},-6.0/{},-5/{t_1},
-4/{},-3/{},-2/{t_2},
-1/{},0/{},1/{t_3},
2/{},3/{},4/{},
5/{t_{4}},6/{},7/{}
}{
\node[v] at (\x,0) {};
\node[left=2pt] at (\x,0) {$\lab$};
}

\draw[box] (-5.5,-7) rectangle (-4,-1.5);
\node[below] at (-4.75,-7) {$L_1$};

\draw[box] (-5.5,1.5) rectangle (-4,7);
\node[above] at (-4.75,7) {$R_1$};

\draw[box] (-2.5,-7) rectangle (-1,-1.5);
\node[below] at (-1.75,-7) {$L_2$};

\draw[box] (-2.5,1.5) rectangle (-1,7);
\node[above] at (-1.75,7) {$R_2$};

\draw[box] (0.5,-7) rectangle (2,-1.5);
\node[below] at (1.25,-7) {$L_3$};

\draw[box] (0.5,7) rectangle (2,1.5);
\node[above] at (1.25,7) {$R_3$};

\draw[box] (4.5,-7) rectangle (6,-1.5);
\node[below] at (5.25,-7) {$L_4$};

\draw[box] (4.5,7) rectangle (6,1.5);
\node[above] at (5.25,7) {$R_4$};

\draw[blue,->]
(-4,-3) to[out=20,in=-120] (-2.5,3);
\draw[blue,->]
(-4,-4) to[out=15,in=-140] (0.5,3);
\draw[blue,->]
(-1,-4) to[out=15,in=-150] (4.5,3);
\draw[blue,->]
(-1,-4) to[out=15,in=-150] (4.5,3);
\draw[blue,->]
(2,-4.5) to[out=20,in=-120] (4.5,4);

\node[circle,fill=black,minimum size=3pt,inner sep=0pt]
at (-5,-6.5) {};
\node[right] at (-5,-6.5) {$t^{in}_{1}$};

\node[circle,fill=black,minimum size=3pt,inner sep=0pt]
at (-5,-5.5) {};

\node[circle,fill=black,minimum size=3pt,inner sep=0pt]
at (-5,-3.5) {};

\node[circle,fill=black,minimum size=3pt,inner sep=0pt]
at (-5,-2.5) {};

\node[circle,fill=black,minimum size=3pt,inner sep=0pt]
at (-5,6.5) {};
\node[right] at (-5,6.5) {$t^{out}_{1}$};

\node[circle,fill=black,minimum size=3pt,inner sep=0pt]
at (-5,5.5) {};

\node[circle,fill=black,minimum size=3pt,inner sep=0pt]
at (-5,3.5) {};

\node[circle,fill=black,minimum size=3pt,inner sep=0pt]
at (-5,2.5) {};

\node[circle,fill=black,minimum size=3pt,inner sep=0pt]
at (-2,-6.5) {};
\node[right] at (-2,-6.5) {$t^{in}_{2}$};

\node[circle,fill=black,minimum size=3pt,inner sep=0pt]
at (-2,-5.5) {};

\node[circle,fill=black,minimum size=3pt,inner sep=0pt]
at (-2,-3.5) {};

\node[circle,fill=black,minimum size=3pt,inner sep=0pt]
at (-2,-2.5) {};

\node[circle,fill=black,minimum size=3pt,inner sep=0pt]
at (-2,6.5) {};
\node[right] at (-2,6.5) {$t^{out}_{2}$};

\node[circle,fill=black,minimum size=3pt,inner sep=0pt]
at (-2,5.5) {};

\node[circle,fill=black,minimum size=3pt,inner sep=0pt]
at (-2,3.5) {};

\node[circle,fill=black,minimum size=3pt,inner sep=0pt]
at (-2,2.5) {};

\node[circle,fill=black,minimum size=3pt,inner sep=0pt]
at (1,-6.5) {};
\node[right] at (1,-6.5) {$t^{in}_{3}$};

\node[circle,fill=black,minimum size=3pt,inner sep=0pt]
at (1,-5.5) {};

\node[circle,fill=black,minimum size=3pt,inner sep=0pt]
at (1,-3.5) {};

\node[circle,fill=black,minimum size=3pt,inner sep=0pt]
at (1,-2.5) {};

\node[circle,fill=black,minimum size=3pt,inner sep=0pt]
at (1,6.5) {};
\node[right] at (1,6.5) {$t^{out}_{3}$};

\node[circle,fill=black,minimum size=3pt,inner sep=0pt]
at (1,5.5) {};

\node[circle,fill=black,minimum size=3pt,inner sep=0pt]
at (1,3.5) {};

\node[circle,fill=black,minimum size=3pt,inner sep=0pt]
at (1,2.5) {};

\node[circle,fill=black,minimum size=3pt,inner sep=0pt]
at (5,-6.5) {};
\node[right] at (5,-6.5) {$t^{in}_{4}$};

\node[circle,fill=black,minimum size=3pt,inner sep=0pt]
at (5,-5.5) {};

\node[circle,fill=black,minimum size=3pt,inner sep=0pt]
at (5,-3.5) {};

\node[circle,fill=black,minimum size=3pt,inner sep=0pt]
at (5,-2.5) {};

\node[circle,fill=black,minimum size=3pt,inner sep=0pt]
at (5,6.5) {};
\node[right] at (5,6.5) {$t^{out}_{4}$};

\node[circle,fill=black,minimum size=3pt,inner sep=0pt]
at (5,5.5) {};

\node[circle,fill=black,minimum size=3pt,inner sep=0pt]
at (5,3.5) {};

\node[circle,fill=black,minimum size=3pt,inner sep=0pt]
at (5,2.5) {};

\draw[red,very thick,midarrow] (-5,-6.5) -- (-5,-5.5);

\draw[red,very thick,midarrow] (-5,-3.5) -- (-5,-2.5);

\draw[red,very thick,midarrow] (-5,-2.5) -- (-5,0);

\draw[red,very thick,midarrow] (-5,0) -- (-5,2.5);

\draw[red,very thick,midarrow] (-5,2.5) -- (-5,3.5);

\draw[red,very thick,midarrow] (-5,5.5) -- (-5,6.5);

\draw[red,very thick,midarrow] (-2,-6.5) -- (-2,-5.5);

\draw[red,very thick,midarrow] (-2,-3.5) -- (-2,-2.5);

\draw[red,very thick,midarrow] (-2,-2.5) -- (-2,0);

\draw[red,very thick,midarrow] (-2,0) -- (-2,2.5);

\draw[red,very thick,midarrow] (-2,2.5) -- (-2,3.5);

\draw[red,very thick,midarrow] (-2,5.5) -- (-2,6.5);

\draw[red,very thick,midarrow] (1,-6.5) -- (1,-5.5);

\draw[red,very thick,midarrow] (1,-3.5) -- (1,-2.5);

\draw[red,very thick,midarrow] (1,-2.5) -- (1,0);

\draw[red,very thick,midarrow] (1,0) -- (1,2.5);

\draw[red,very thick,midarrow] (1,2.5) -- (1,3.5);

\draw[red,very thick,midarrow] (1,5.5) -- (1,6.5);

\draw[red,very thick,midarrow] (5,-6.5) -- (5,-5.5);

\draw[red,very thick,midarrow] (5,-3.5) -- (5,-2.5);

\draw[red,very thick,midarrow] (5,-2.5) -- (5,0);

\draw[red,very thick,midarrow] (5,0) -- (5,2.5);

\draw[red,very thick,midarrow] (5,2.5) -- (5,3.5);

\draw[red,very thick,midarrow] (5,5.5) -- (5,6.5);


\draw[dotted, thick] (-5,-5.5) -- (-5,-3.5);
\draw[dotted, thick] (-5,5.5) -- (-5,3.5);

\draw[dotted, thick] (-2,-5.5) -- (-2,-3.5);
\draw[dotted, thick] (-2,5.5) -- (-2,3.5);

\draw[dotted, thick] (1,-5.5) -- (1,-3.5);
\draw[dotted, thick] (1,5.5) -- (1,3.5);

\draw[dotted, thick] (5,-5.5) -- (5,-3.5);
\draw[dotted, thick] (5,5.5) -- (5,3.5);


\node[draw,thick,align=left] at (8,7) {
$(t_1,t_4)\in \mathcal{T}$\\
$(t_2,t_3)\in \mathcal{T}$
};

\end{tikzpicture}
\caption{Schematic illustration of the construction for the directed variant. 
For each terminal $t_i$, the construction adds a lower chain $L_i$ entering $t_i$ and an upper chain $R_i$ leaving $t_i$. 
Shortcut arcs are added from $L_i$ to $R_j$ for non-terminal pairs. The red arcs denote reinforced paths of length two.}
\label{fig:directed-construction}
\end{figure}

Finally, we set $\ell := 2m$ where $m=|A(G)|$, and define $s := 4\ell$ and $k':=k$.
All arcs thus point from smaller to larger terminal indices, or stay within one gadget
in the forward direction $t_i^{\mathrm{in}}\to t_i\to t_i^{\mathrm{out}}$.

It is straightforward to verify that $D'$ remains acyclic:
every new arc is directed from a vertex associated with $t_i$
towards vertices associated with $t_j$ for some $j \ge i$,
and all reinforced or cross-connection arcs respect this ordering.
Hence no directed cycle is introduced by the construction.

\proofpara{Forward direction (\texorpdfstring{$\Rightarrow$}{⇒})}
Assume there exists a multicut $F\subseteq A(G)$ with $|F|\le k$
such that for every $(t_i,t_j)\in\mathcal{T}$,
there is no directed path from $t_i$ to $t_j$ in $G-F$.
We claim that deleting the same set of arcs $F$ from $D'$
yields a digraph in which every weakly connected component is $s$-reach-bounded
with $s=4\ell$.
That is, for all vertices $u,v$ in the same weakly connected component of $D'-F$,
if $v$ is reachable from $u$ in $D'-F$, then $\mathrm{dist}_{D'-F}(u,v)\le s$.

We consider cases depending on the position of $u$ and $v$.

\smallskip
\noindent
\textbf{Case~A.} \emph{$u$ and $v$ lie within the same terminal gadget.}
Each gadget corresponding to $t_i$ consists of two reinforced chains
$t_i^{\mathrm{in}}\!\to t_i$ and $t_i\!\to t_i^{\mathrm{out}}$.
There are no shortcut arcs from $L_i$ to $R_i$ inside the same gadget.
Each reinforced step contributes length~2, and there are at most $\ell$ such
steps on each of the two chains. Hence any directed path inside the gadget
has length at most $4\ell=s$.
Thus, whenever $v$ is reachable from $u$ within the same gadget,
\[
\mathrm{dist}_{D'-F}(u,v)\le 4\ell=s.
\]

\smallskip
\noindent
\textbf{Case~B.} \emph{$u$ and $v$ lie in gadgets $i$ and $j$ with $(t_i,t_j)\notin\mathcal{T}$.}
By construction, for every pair $(t_i,t_j)\notin\mathcal{T}$ with $i<j$,
we added all arcs from vertices of $L_i$ to vertices of $R_j$.
We analyze according to the position of $u$ inside gadget $i$.

\emph{B.1. $u\in L_i$.}
Then $u$ has a direct outgoing arc to every $y\in R_j$.
If $v\in R_j$, we can choose $y=v$ and obtain
$\dist_{D'-F}(u,v)=1\le s$.
If $v\in L_j$, then $u$ does not reach $v$, since every arc inside gadget $j$
is directed towards $t_j$ along $L_j$ and away from $t_j$ along $R_j$,
and there are no arcs from $R_j$ back to $L_j$.
Hence the condition holds vacuously.

\emph{B.2. $u\in R_i$.}
No vertex of $R_i$ has outgoing arcs to vertices of other gadgets.
Thus $u$ cannot reach any vertex in gadget $j$, and the condition again holds vacuously.

\emph{B.3. $u=t_i$.}
There is no direct shortcut arc from $t_i$ to gadget $j$.
However, if $t_i$ reaches $t_j$ in the original DAG $G$,
then there exists a directed $t_i\to t_j$ path in $G$ of length at most $m$.
From $t_j$, every vertex of $R_j$ is reachable within at most $2\ell$ steps
along the reinforced chain from $t_j$ to $t_j^{\mathrm{out}}$.
Hence, whenever $t_i$ reaches some $v\in R_j$, we obtain
\[
\dist_{D'-F}(t_i,v)
\le m+2\ell
\le 5m
\le 8m
=4\ell
=s,
\]
since $\ell=2m$ and $s=4\ell$.
If $v\in L_j$, then $t_i$ cannot reach $v$, because there are no arcs from
$R_j$ to $L_j$. Hence the condition holds vacuously.

\smallskip
\noindent
\textbf{Case~C.} \emph{$u$ and $v$ lie in gadgets $i$ and $j$ with $(t_i,t_j)\in\mathcal{T}$.}
For terminal pairs $(t_i,t_j)\in\mathcal{T}$, no shortcut arcs from
$L_i$ to $R_j$ were added in the construction.
Hence any directed path from gadget $i$ to gadget $j$
must correspond to a directed path from $t_i$ to $t_j$ in the original DAG $G$.

However, since $F$ is a valid multicut for the instance
$(G,\mathcal{T},k)$, the graph $G-F$ contains no directed path from
$t_i$ to $t_j$.
Consequently, after deleting the arcs of $F$ from $D'$,
no vertex of gadget $i$ can reach any vertex of gadget $j$ in $D'-F$.
Therefore, the $s$-reach-bounded property holds vacuously in this case.

In all cases, if $v$ is reachable from $u$ in $D'-F$,
then $\mathrm{dist}_{D'-F}(u,v)\le s$.
Hence, every weakly connected component of $D'-F$
is $s$-reach-bounded with $s=4\ell$,
as required.

\proofpara{Backward direction}
Assume that $F\subseteq A(D')$ is a feasible solution for 
\textsc{$s$-Club Cluster Arc Deletion} on $D'$ with $|F|\le k$,
and let $F_G := F \cap A(G)$ denote the corresponding set of arcs removed from the original DAG~$D$.
Suppose, for contradiction, that there exists a terminal pair 
$(t_i,t_j)\in\mathcal{T}$ such that a directed path from $t_i$ to $t_j$
remains in $G-F_G$.

In $D'-F$, the following hold:
\begin{enumerate}
    \item The reinforced path from $t_i^{\mathrm{in}}$ to $t_i$ has length exactly $2\ell$
    and remains intact, since every replaced arc was substituted by $k+1$
    internally vertex-disjoint directed 2-paths, and $|F|\le k$.
    \item Similarly, the path from $t_j$ to $t_j^{\mathrm{out}}$ also has length $2\ell$
    and is unaffected by the deletion of $F$.
    \item No shortcut arcs were added from $L_i$ to $R_j$ for $(t_i,t_j)\in\mathcal{T}$,
    so any path from $t_i^{\mathrm{in}}$ to $t_j^{\mathrm{out}}$ must pass through
    $t_i$, traverse a copy of the original $t_i\!\to t_j$ path in $G-F_G$,
    and then proceed through the reinforced segment leading to $t_j^{\mathrm{out}}$.
\end{enumerate}

Hence, concatenating these parts yields a directed path in $D'-F$ as
~$t_i^{\mathrm{in}} \;\leadsto\; t_i \;\leadsto\; t_j \;\leadsto\; t_j^{\mathrm{out}}$,
whose total length is at least
$2\ell \;+\; 1 \;+\; 2\ell \;=\; 4\ell + 1$.
The first and last $2\ell$ correspond to the reinforced segments to and from the terminals,
while the middle part contributes at least one original arc of $G$.
Since $s = 4\ell$, we have
$\mathrm{dist}_{D'-F}\bigl(t_i^{\mathrm{in}},t_j^{\mathrm{out}}\bigr) \ge 4\ell + 1 > s$.
Therefore, vertices $t_i^{\mathrm{in}}$ and $t_j^{\mathrm{out}}$ lie in the same weakly connected component of $D'-F$
and violate the $s$-reach-bounded property,
contradicting the assumption that $F$ is a valid solution.
It follows that no terminal pair $(t_i,t_j)\in\mathcal{T}$ remains reachable in $G-F_G$,
and thus $F_G$ is a valid multicut for $(G,\mathcal{T},k)$.
\end{proof}

\section{Conclusion and Open Problems}

This work presented a systematic study of the parameterized and structural complexity of the \textsc{$s$-Club Cluster Edge Deletion} problem,  
a natural generalization of \textsc{Cluster Edge Deletion} in which cliques are replaced by $s$-clubs.  
Our investigation was guided by a central question:  
\emph{when is the diameter bound $s$ essential for achieving tractability?}  
The results reported here delineate the fine boundary between fixed-parameter tractability and hardness,  
showing how the interplay between structural parameters and distance constraints governs computational feasibility.

\medskip
\noindent\textbf{Parameterized complexity.}  
From the viewpoint of parameterized complexity,  
we established that the problem is fixed-parameter tractable  
when parameterized by neighborhood diversity, twin cover, and cluster vertex deletion number.  
These results extend to all $s\ge1$ the FPT algorithms previously known only for the case $s=1$,  
thus unifying and generalizing several earlier approaches.  
On the hardness side, we proved that the problem is W[1]-hard when parameterized by pathwidth—and consequently by treewidth—  
thereby resolving an open question posed by Montecchiani et al.\ and demonstrating that the dependence on $s$ in their FPT algorithm is indeed unavoidable.
In fact, our construction yields a stronger result:  
{\sc $s$-Club Cluster Edge Deletion} remains W[1]-hard even when parameterized by the combined parameter  
$\pw(G) + d$, where $d$ denotes the maximum number of $s$-clubs allowed in the resulting graph.

\medskip
\noindent\textbf{Bicriteria approximation and the role of the diameter bound.}  
To further probe the influence of the diameter constraint,  
we studied the single-parameter regime where only the solution size $k$ is bounded.  
While the existence of an exact FPT algorithm parameterized solely by $k$ remains open,  
we made progress by designing a bicriteria fixed-parameter approximation algorithm.  
For graphs excluding long induced cycles, our algorithm runs in time  
$f(k,1/\varepsilon)\,n^{\mathcal{O}(1)}$  
and outputs a solution of size at most~$k$ whose connected components have diameter at most $(1+\varepsilon)s$.  
This shows that a small relaxation of the diameter bound can restore fixed-parameter tractability,  
highlighting the subtle and central role played by distance constraints in defining computational hardness.

\medskip
\noindent\textbf{Directed variant.}
We additionally study the directed variant of
\textsc{$s$-Club Cluster Edge Deletion} parameterized by the cutset size~$k$,
that is, by the number of arcs allowed to be deleted.
We show that this directed variant is W[1]-hard even when the input
graph is a directed acyclic graph (DAG).

\medskip
\noindent\textbf{Future directions.}  
Together, these contributions nearly complete the parameterized landscape of  
\textsc{$s$-Club Cluster Edge Deletion} under structural measures,  
while opening several compelling directions for future research.  
The most immediate challenge is to determine whether the problem admits an \emph{exact} FPT algorithm when parameterized solely by $k$.  
Other promising directions include exploring parameterizations by feedback vertex set, modular width, or feedback edge set,  
and investigating the existence of polynomial kernels under the combined parameter $(s+k)$—  
a question that remains open~\cite{MISRA2020150}.

\medskip
In summary, our results pinpoint when and why the diameter bound~$s$ becomes crucial for tractability.  
By linking local distance constraints with global structural graph properties,  
this work advances the theoretical understanding of how bounded-diameter clustering problems transition  
from efficiently solvable to intractable regimes.

\bibliographystyle{elsarticle-num} 
\bibliography{Referances}

\end{document}